\documentclass[aps,prx,twocolumn,showpacs]{revtex4-1}
\pdfoutput=1
\usepackage{hyperref}
\hypersetup{
colorlinks = true, linkcolor = [rgb]{0,0,1}, citecolor = [rgb]{0,0,1},
                             urlcolor = [rgb]{0,0,1}}
\usepackage{graphicx,epsf}
\usepackage{amsmath}
\usepackage{amssymb}
\usepackage{color}
\usepackage{esint}
\usepackage{wasysym}
\newcommand{\bq}{{\bf q}}

\newcommand{\bp}{{\bf p}}
\newcommand{\bk}{{\bf k}}

\newcommand{\taub}{\mbox{\boldmath $\tau $}}
\newcommand{\deltab}{\mbox{\boldmath $\delta $}}

\begin{document}

\title{Flat-band ferromagnetism in a correlated topological insulator on a
         honeycomb lattice}

\author{Leonardo S. G. Leite  and R. L. Doretto}
\affiliation{Instituto de F\'isica Gleb Wataghin,
                  Universidade Estadual de Campinas,
                  13083-859 Campinas, SP, Brazil}

\date{\today}

\begin{abstract}

We study the flat-band ferromagnetic phase of a spinfull and
time-reversal symmetric Haldane-Hubbard model on a honeycomb lattice
within a bosonization formalism for flat-band Z$_2$ topological insulators. 
Such a study extend our previous one [Phys. Rev. B {\bf 104}, 155129 (2021)] 
concerning the flat-band ferromagnetic phase of a correlated Chern
insulator described by a Haldane-Hubbard model.
We consider the topological Hubbard model at $1/4$ filling of its
corresponding noninteracting limit and in the nearly flat band limit
of its lower free-electronic bands. We define boson operators associated with
two distinct spin-flip excitations, one that changes (mixed-lattice excitations) and 
a second one that preserves (same-lattice excitations) 
the index related with the two triangular sublattices.  
Within the bosonization scheme, the fermion model is mapped
into an effective interacting boson model, whose quadratic term is
considered at the harmonic approximation in order to determine the
spin-wave spectrum.
For both mixed- and same-lattice excitations, we find that the spin-wave
spectrum is gapped and has two branches, with an
energy gap between the lower and the upper bands at the $K$ and $K'$
points of the first Brillouin zone.
We find that the same-lattice excitations are indeed the lowest-energy
(elementary) excitations that characterize the flat-band ferromagnetic phase, a
feature that contrasts with the behaviour of a previously studied 
correlated topological insulator on a square lattice,
whose flat-band ferromagnetic phase is characterized by mixed-lattice excitations.
We also find some evidences that the spin-wave bands for the
same-lattice excitations might be topologically nontrivial even in the
completely flat band limit.

\end{abstract}
%\pacs{xxx}

\maketitle

%%%%%%%%%%%%%%%%%%%%%%%%%%%%%%%%%%%%%%%%%%%%%%%%%%%%%%%%%%%%%%%%%%%%%%%%%%%%%%%%%%%%%
\section{Introduction}
\label{sec:intro}

The first theoretical proposal of a Chern band insulator
came from a pioneering work of Haldane in 1988 \cite{haldane1988model}. 
In that paper, Haldane introduced a spinless tight-binding model on a
honeycomb lattice with broken time-reversal symmetry that even without
an external source of magnetic field displays a quantum Hall effect.  
The emergence of this distinct insulating quantum Hall phase derives
from the topologically nontrivial electronic band structure of the
Haldane model: the nonzero Chern numbers \cite{thouless1982} 
of these electronic bands yield a finite Hall conductivity at half-filling, i.e.,
the system exhibits the so-called anomalous quantum Hall effect
\cite{qi11,review-aqhe}.

The Haldane model on a honeycomb lattice was later geneneralized by 
Kane and Mele \cite{kane-mele05,kane2005quantum}, 
providing the first microscopic model for  a topological insulator
\cite{hasan2010colloquium,kane13}. 
Here the spin degree of freedom is explicitly included and, in contrast with
Haldane model, time-reversal symmetry is preserved.
Although at half filling time-reversval symmetry yields a vanishing
total Chern number, such a system may exhibit a quantum spin Hall
effect \cite{kane-mele05,kane2005quantum,zhang06}. 
Indeed, the Kane-Mele model is an example of 
a Z$_2$ topological insulator, a system which is characterized by a Z$_2$
invariant that distinguishes between the trivial insulator phase and
the topologically nontrivial one \cite{hasan2010colloquium,kane13}.    
In spite of the fact that the Kane-Mele model 
is not experimentally realized so far, the quantum spin Hall effect 
was theoretically predicted \cite{bernevig2006quantum}   
and later experimentally observed \cite{konig2007quantum} 
in HgTe/CdTe quantum wells at low temperatures. 
Interestingly, experimental implementations of topological
insulators using ultracold atoms in optical lattices have also been considered 
\cite{zoller16,review-adv-phys18,rmp-cooper19}.

Correlation effects on topological insulators have also been
receiving some attention in recent years \cite{rachel2018, hohenadler2013}.  
An interesting example of a correlated topological insulator on a
honeycomb lattice is the Kane-Mele-Hubbard model
\cite{rachel2010, assad2011,xie2011,zheng2011,hohenadler2012,hung13,lang13,hung14,laubach14,klein21},
which is a generalization of the Kane-Mele model with the
electron-electron interaction being described by an on-site Hubbard repulsion term. 
The phase diagram of the  
model has been determined \cite{rachel2010,zheng2011,hohenadler2012} at half filling. 
In particular, quantum Monte Carlo simulations have been
performed \cite{zheng2011,hohenadler2012}, 
since, in this case, the so-called fermion sign problem is absent, a
feature that is related to the fact that the model preserves
particle-hole symmetry at half filling \cite{zheng2011}.
It was shown that, apart from some possible intermediate phases,  
a $Z_2$ topological band insulator phase survives
for small to moderate values of  the on-site repulsion energy $U$
and that the system enters a magnetically ordered phase   
above a critical on-site repulsion energy $U_c$.
An analytical description of such Mott transition was recently performed 
\cite{klein21}.

Another set of interacting topological systems that has been recently
gaining some attention is made out of lattice models that display (nearly)
flat and topologically nontrivial electronic bands in the noninteracting limit
\cite{katsura2010,neupert2012topological,doretto2015flat,su2018topological,
        su2019ferromagnetism,gu2019itinerant,gu21,leite2021}. 
In a sense, these papers transport the long discussed subject of
flat-band ferromagnetism
\cite{kusakabe1994,tasaki1996,flat-fm} to the realm of lattice models
with topologically nontrivial free-electronic bands.
Indeed, the merging of these two subjects was motivated by a series of
papers \cite{neupert2011fractional, sun2011nearly,tang2011high} that
describe tight-binding models, specially in two dimensions, with only
short-range hoppings and whose parameters, once fine tuned, may yield
nearly flat and topologically nontrivial electronic bands. 
In particular, in Ref.~\cite{leite2021}, we studied the flat-band
ferromagnetic (FM) phase of a correlated Chern insulator on a honeycomb
lattice described by a Haldane-Hubbard model.
We considered the model at $1/4$ filling (half filling of the lower
and doubly degenerated free-electronic band) and in the vicinity of
a suitable choice of the model parameters
\cite{neupert2011fractional}, that yields nearly flat
noninteracting bands.
In order to describe such a flat-band FM phase, we employed 
a bosonization scheme for flat-band correlated Chern insulators \cite{doretto2015flat}, that
was developed by one of us.   
Such a formalism allows us to map the Haldane-Hubbard model 
to an effective interacting boson model:
We considered the effective boson model within a harmonic approximation 
and determined the spin-wave spectrum;
it was found that the excitation spectrum has one gapped and one
gapless excitation branches, with a Goldstone mode at the center of
the first Brillouin zone (a feature that indicates the stability of
the flat-band FM phase) and Dirac points 
at the $K$ and $K'$ points of the first Brillouin zone (BZ).

In the present paper, we extend our previous study \cite{leite2021}
about the flat-band FM phase of a correlated Chern
insulator on a honeycomb lattice, by considering a similar,
but now time-reversal symmetric, topological
Hubbard (THM) model on a honeycomb lattice.
The noninteracting term of such correlated Z$_2$ topological insulator
is given by a spinfull version of the Haldane model \cite{haldane1988model} 
that preserves time-reversal symmetry. 
Similarly to Ref.~\cite{leite2021}, we consider the THM
at $1/4$ filling of its noninteracting limit and in the
vicinity of the nearly flat band limit \cite{neupert2011fractional} 
of its lower free-electronic band.
The flat-band FM phase of the time-reversal symmetric
THM is described within a bosonization scheme for
flat-band correlated Z$_2$ topological insulators, a formalism that was introduced
in Ref.~\cite{doretto2015flat} and is based on the bosonization formalism \cite{doretto2005}
proposed to study the quantum Hall system at filling factor $\nu = 1$.
Again, the THM is mapped to an effective
interacting boson model. We define boson operators
[Eq.~\eqref{eq:bosons}] associated with two distinct spin-flip 
excitations that are termed 
mixed-lattice [Eq.~\eqref{eq:projS1}] and
same-lattice [Eq.~\eqref{eq:projS2}] excitations.
In both cases, we find that the spin-wave excitation spectrum is
gapped and constituted by two bands completely separated from each
other, a feature that contrasts with the spin-wave spectrum of the correlated Chern
insulator \cite{leite2021}, whose bands touch at the corners of the first BZ.  
Interestingly, in contrast with the square lattice correlated
topological insulator \cite{doretto2015flat},
whose flat-band FM phase is characterized by mixed-lattice excitations,
here, for the correlated topological insulator on a honeycomb lattice,
we find that the same-lattice ones are indeed the correct mode, which
furnishes the lowest-energy excitations [see Figs.~\ref{figEspectro}(a)-(f)].
Finally, we also find some indications that the spin-wave excitation
bands for the same-lattice excitations might be topologically
nontrivial, since the corresponding Chern numbers are nonzero.
As far as we know, this is the first calculation of the spin-wave spectrum
for the flat-band FM phase of a correlated $Z_2$ topological
insulator on a honeycomb lattice described by a Haldane-Hubbard like
model.

Our paper is organized as follows. 
In Sec.~\ref{sec:TBmodel}, we introduce the time-reversal symmetric
THM on a honeycomb lattice.
In Sec.~\ref{sec:boso}, the bosonization formalism for flat-band Z$_2$
topological insulators \cite{doretto2015flat} is briefly reviewed.
In Sec.~\ref{sec:flatferromagnetism},  the effective interacting boson
model, that allows us to described the flat-band FM 
phase of the correlated topological insulator, is presented;  
the boson model is considered within the harmonic approximation:
the spin-wave spectrum is determined for homogeneous and
sublattice dependent on-site Hubbard repulsion energies.
Section~\ref{sec:summary} contains a brief summary of our main results. 
Some details of the bosonization formalism and additional results are
presented in the five Appendices.

\begin{figure*}[t]
\centerline{ 
 \includegraphics[width=4.5cm]{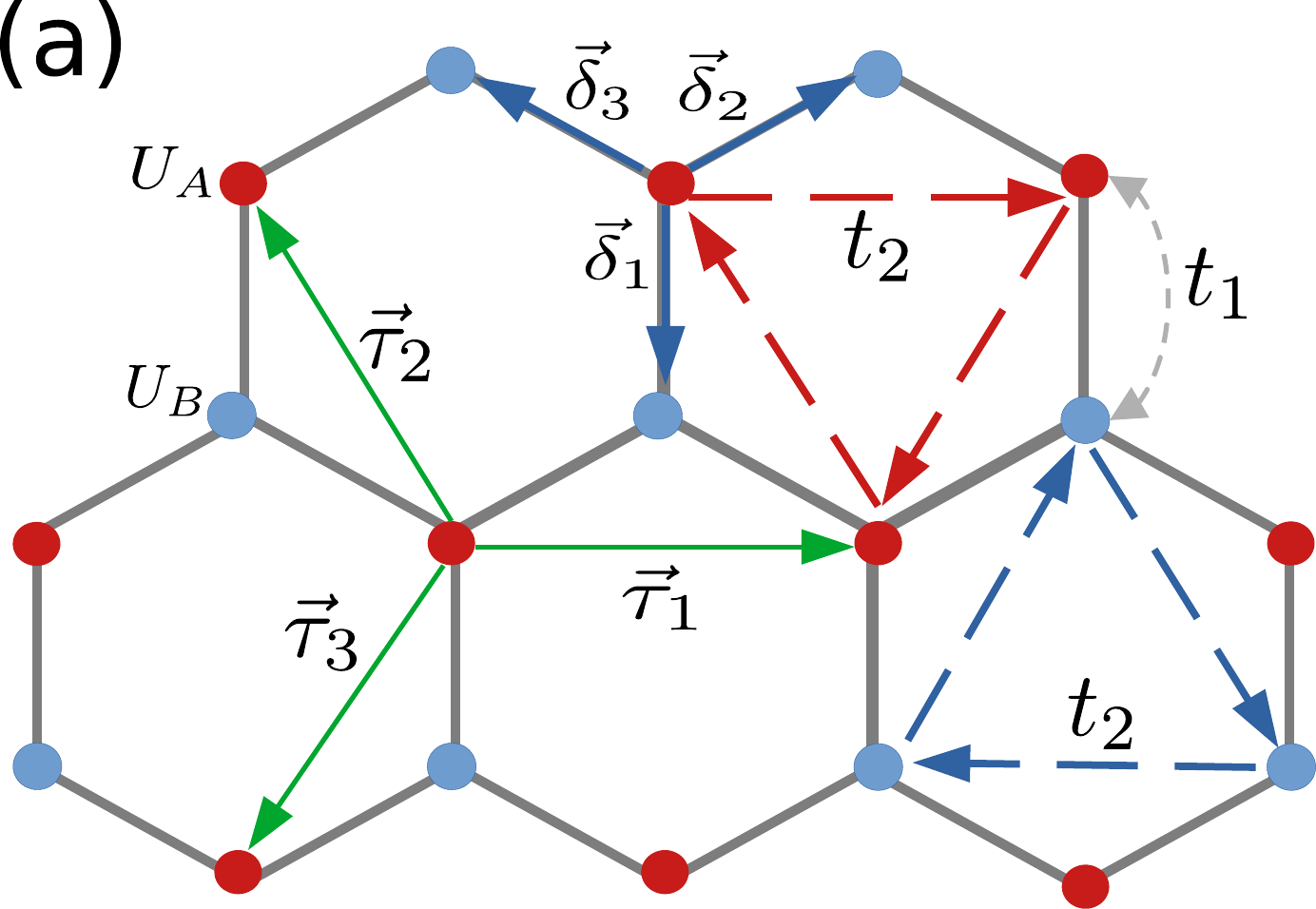} 
 \hskip1.5cm
 \includegraphics[width=3.0cm]{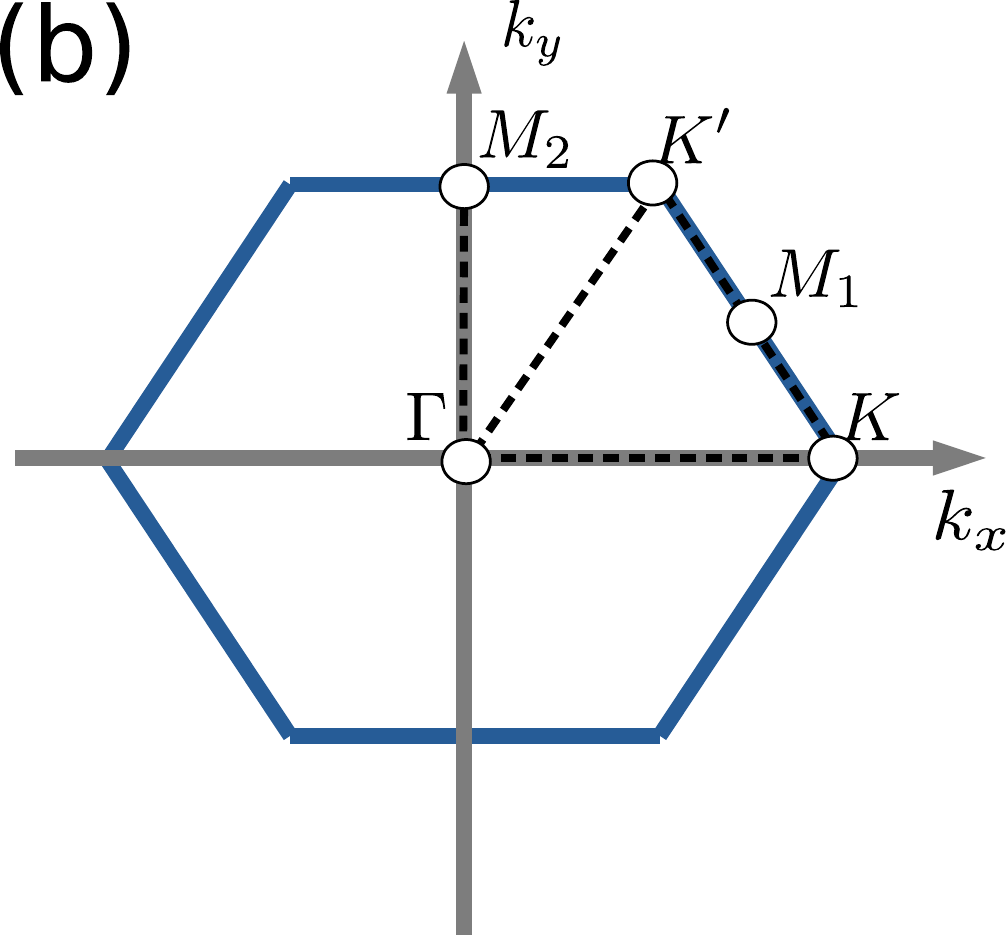}
 \hskip1.5cm
 \includegraphics[width=5.0cm]{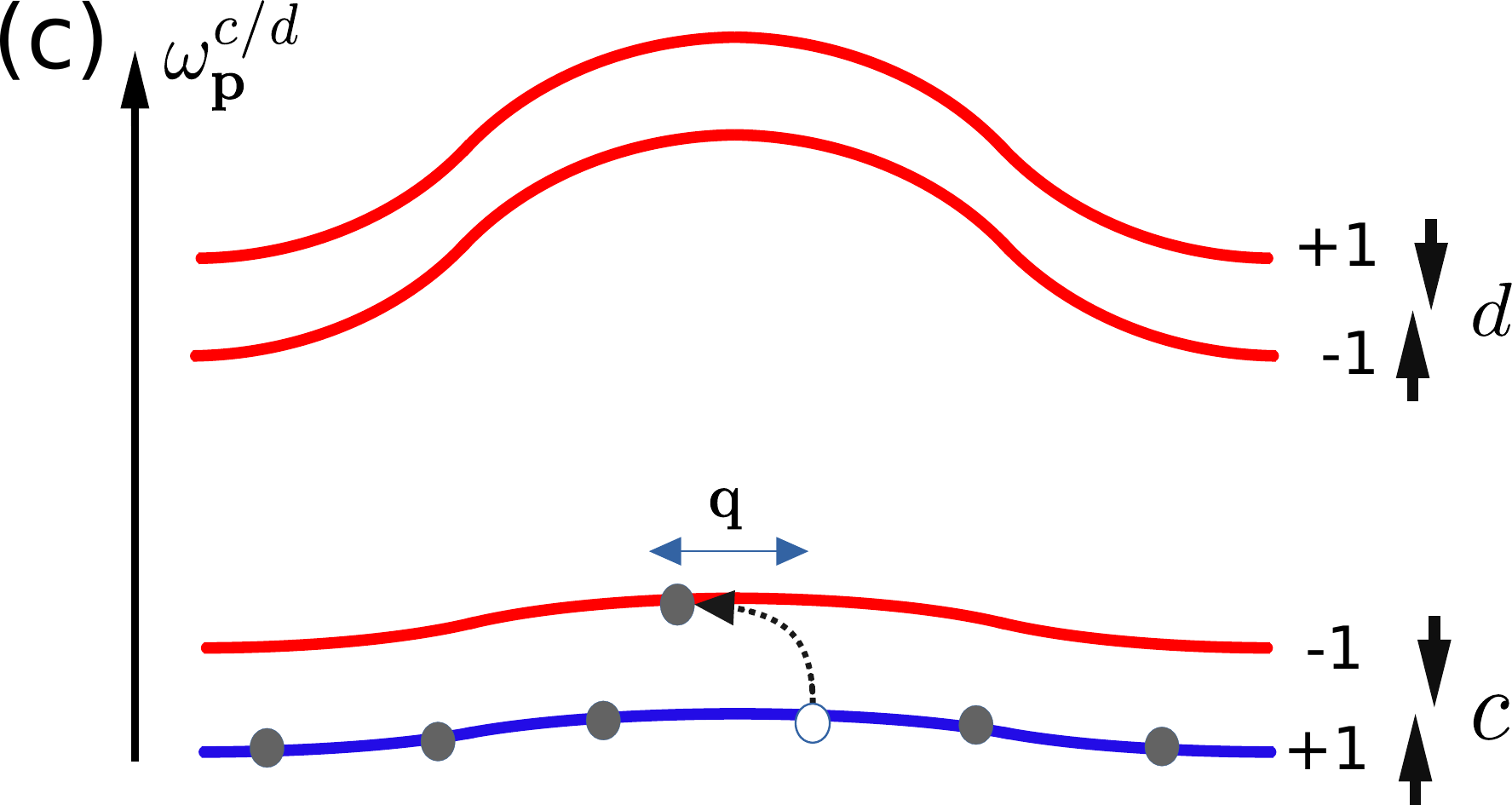}
}
\caption{
(a) Schematic representation of the THM 
model \eqref{eqHH} on an honeycomb lattice. Red and blue circles 
respectively represent the sites of the (triangular) sublattices $A$ and $B$. 
The nearest-neighbor and next-nearest-neighbor hopping
energies are given by $t_1$ and $t_2 e^{\pm i \phi}$
(positive sign follows arrow direction), respectively,
while $U_A$ and $U_B$ indicate the sublattice dependent on-site
Hubbard repulsion energies.  
The nearest-neighbor \eqref{deltavectors}
and next-nearest-neighbor \eqref{tauvectors} vectors are indicate by 
$\deltab_i$ and $\taub_i$, respectively.
(b) The first BZ, where   
$\mathbf{K} = ( 4\pi/3\sqrt{3}, 0 )$,
$\mathbf{K'} = ( 2\pi/3\sqrt{3}, 2\pi/3 )$,
$\mathbf{M}_1 = ( \pi/\sqrt{3}, \pi/3)$, and
$\mathbf{M}_2 = ( 0, 2\pi/3 )$,
with the nearest-neighbor distance of the honeycomb lattice $a = 1$.
(c) Schematic representation of the noninteracting electronic bands
\eqref{eq:omega} in the nearly-flat band limit \eqref{optimal-par} of
the lower bands $c$. At $1/4$ filling, the ground state is the
FM state \eqref{eq:FM} and low energy excitations are
particle-hole pairs (spin flips) within the lower bands. 
Although the noninteracting bands $c$ and $d$ are doubly degenerated
with respect to the spin degree of freedom, we introduce an offset between the    
$\sigma = \uparrow$ and $\downarrow$ bands for clarity.
The Chern numbers \eqref{eqCn} of each band are
also shown on the right side.}
\label{fig:Lattice}
\end{figure*}

%%%%%%%%%%%%%%%%%%%%%%%%%%%%%%%%%%%%%%%%%%%%%%%%%%%%%%%%%%%%%%%%%%%%%%%%%%%%%%%%%%%%%
\section{The time-reversal symmetric Haldane-Hubbard model}
\label{sec:TBmodel}

In this section, we introduce a time-reversal symmetric Haldane-Hubbard
model on a honeycomb lattice. 
Our discussion closely follows the lines of Sec.~II from Ref.~\cite{leite2021},
where such a Haldane-Hubbard model with broken time-reversal symmetry
is described.

%%%%%%%%%%%%%%%%%%%%%%%%%%%%%%%%%%%%%%%%%%%%%%%%%%%%%%%%%%%%%%%%%%%%%%%%%%%%%%%%%%%%%
\subsection{The fermionic interacting model}
\label{sec:model}

Let us consider $N_e$ spin-$1/2$ electrons on a honeycomb lattice 
described by a Haldane-Hubbard model, whose Hamiltonian assumes the
form 
\begin{equation}
    H =  H_0 + H_U, 
\label{eqHH}
\end{equation}
where the noninteracting term is given by
\begin{align}
 H_0 &=  t_1 \sum_{i \in A, \delta, \sigma}  \left( c_{i A \sigma}^{\dagger} c_{i + \delta B \sigma}  
                      + {\rm H.c.} \right) 
\nonumber \\
	&+ t_2  \sum_{i \in A, \tau, \sigma}   
                 \left( e^{-i\gamma_\sigma\phi} c_{i A \sigma}^{\dagger} c_{i + \tau  A \sigma}   + {\rm H.c.} \right) 
\nonumber \\
        &+ t_2  \sum_{i \in B, \tau, \sigma}   
                 \left( e^{+i\gamma_\sigma\phi} c_{i B \sigma}^{\dagger} c_{i + \tau  B \sigma}   + {\rm H.c.} \right),
\label{eqHH0}
\end{align}
while the interacting one is an on-site Hubbard repulsion term,
\begin{equation}
   H_U = \sum_i \sum_{a = A,B} U_a \hat{\rho}_{i a \uparrow} \hat{\rho}_{i a \downarrow}.
\label{eqHHU}
\end{equation}
Here the operator $c_{i a \sigma}^{\dagger}$  ($c_{i a \sigma}$) creates (destroys) 
an electron with spin $\sigma = \uparrow, \downarrow$ on
the $i$-th site of the (triangular) sublattice $a = A$, $B$ of the
honeycomb lattice.  
The nearest-neighbor and next-nearest-neighbor hopping amplitudes  
are both positive and given by $t_1$ and $t_2$, respectively
[see Fig.~\ref{fig:Lattice}(a)]. 
Indeed, the next-nearest-neibhbor hopping is complex, 
$t_2e^{\pm i \gamma_\sigma\phi}$, which indicates that the electron 
acquires a (spin-dependent) $+\gamma_\sigma\phi$ phase
and a $-\gamma_\sigma\phi$ phase as it moves, respectively, 
in the same and opposite directions of the arrows associated with the
dashed lines in Fig.~\ref{fig:Lattice}(a)   
(see also note \cite{comment01}).
The complex next-nearest-neibhbor hopping yields a fictitious flux
pattern with zero net flux per unit cell \cite{neupert2011fractional}. 
Importantly, time-reversal invariance requires that 
$\gamma_\uparrow = - \gamma_\downarrow = 1$,
which implies that the spin $\uparrow$ electrons and the 
spin $\downarrow$ electrons experience an opposite fictitious flux
pattern (see also Sec.~II from Ref.~\cite{doretto2015flat}). 
The index $\delta$ indicates the nearest-neighbor vectors
\begin{align}
 \deltab_1 &= -a\hat{y}, 
\quad\quad
 \deltab_{2,3} = \pm\frac{a}{2}\left( \sqrt{3}\hat{x} \pm \hat{y} \right), 
\label{deltavectors}
\end{align}
as illustrated in Fig.~\ref{fig:Lattice}(a),
and $\tau$ corresponds to the next-nearest-neighbor vectors
$\taub_1 = \deltab_2 - \deltab_3$, 
$\taub_2 = \deltab_3 - \deltab_1$, and 
$\taub_3 = \deltab_1 - \deltab_2$:
\begin{align}
 \taub_1 &= a\sqrt{3}\hat{x}, 
\quad\quad
 \taub_{2,3} = -\frac{a}{2}\left( \sqrt{3}\hat{x} \mp 3\hat{y} \right).
\label{tauvectors}
\end{align}
Hereafter, we set the nearest-neigbhor distance $a = 1$.  
One should mention that, for $\phi = \pi/2$, 
the tight-binding model \eqref{eqHH0} corresponds to the
Kane-Mele model in the absence of the Rashba term \cite{kane-mele05}.
Finally, $\hat{\rho}_{i a\sigma}$ is the density operator for 
spin $\sigma$ electrons at site $i$ of sublattice $a$,
\begin{equation}
   \hat{\rho}_{i a\sigma} = c_{i a \sigma}^{\dagger} c_{i a \sigma},
\label{dens-op}
\end{equation}
and $U_a > 0$ are the on-site and sublattice-dependent repulsion
energies.

%%%%%%%%%%%%%%%%%%%%%%%%%%%%%%%%%%%%%%%%%%%%%%%%%%%%%%%%%%%%%%%%%%%%%%%%%%%%%%%%%%%%%
\subsection{Diagonalization of the noninteracting Hamiltonian}
\label{sec:Diagonalization}

In order to diagonalize the noninteracting model \eqref{eqHH0}, 
one considers the Fourier transform
\begin{equation}
  c_{i a \sigma}^{\dagger} = \frac{1}{\sqrt{N_a}} \sum_{\bk \in {\rm BZ}} 
                  e^{i \mathbf{k} \cdot \mathbf{R}_i} c_{ \mathbf{k} \, a \, \sigma}^{\dagger} ,
\label{eq:Fourier}
\end{equation}
where $N_a = N$ is the number of sites of the sublattice $a$
and the momentum sum runs over the first BZ
associated with the underline triangular Bravais lattice, 
see Fig.~\ref{fig:Lattice}(b).
The noninteracting Hamiltonian \eqref{eqHH0} can then be written in a
matrix form, i.e.,  
\begin{equation}
  H_{0} = \sum_{\mathbf{k}}  \Psi_{\mathbf{k} }^{\dagger}  H_\bk \Psi_{\mathbf{k} },  
\label{eqH0k}
\end{equation}	
where the $4 \times 4$ $H_\bk$ matrix reads
\begin{equation}
  H_\bk =  \left(\begin{array}{cc}
			h^{\uparrow}_{\mathbf{k}} &  0 \\ 
			0  &  h^{\downarrow}_{\mathbf{k}}
			\end{array}  \right)
			\label{Hmatrix}
\end{equation}
and the four-component spinor $\Psi_{\mathbf{k}} $ is defined as  
\begin{equation}
\Psi_{\mathbf{k}} = \left( 
			c_{ \mathbf{k} A \uparrow} \;\; 
			c_{ \mathbf{k} B \uparrow} \;\;
			c_{ \mathbf{k} A \downarrow} \;\; 
			c_{ \mathbf{k} B \downarrow} \right)^T.
\end{equation}
The $2 \times 2$ matrices $h^\uparrow_\bk$ and $h^\downarrow_\bk$
associated with each spin sector in Eq.~\eqref{Hmatrix} can be written
in terms of the identity matrix $\tau_0$ and the vector 
$\hat{\tau} = (\tau_1$, $\tau_2$, $\tau_3)$, whose 
components are Pauli matrices, i.e.,
\begin{equation}
    h_\bk^\sigma  = B_{0, \mathbf{k} }^\sigma  \tau_0 +  \mathbf{B}_{\mathbf{k} }^\sigma  \cdot \hat{\tau} ,
\end{equation}	
where $\mathbf{B}^\sigma_{\mathbf{k} } = ( B^\sigma_{1,\mathbf{k} }, B^\sigma_{2, \mathbf{k} }, B^\sigma_{3, \mathbf{k} })$ and 
\begin{align}
  B^\sigma_{0,\mathbf{k} }  &= B_{0,\mathbf{k} }  
               = 2 t_2 \cos(\phi)\sum_{\mathbf{\tau}} \cos(\mathbf{k} \cdot \taub ) , 
\nonumber \\
  B^\sigma_{1,\mathbf{k} } &= B_{1,\mathbf{k} }  
               = t_1  \sum_{\mathbf{\delta}} \cos(\mathbf{k} \cdot \deltab) ,
\nonumber \\ 
  B^\sigma_{2,\mathbf{k} } &= B_{2,\mathbf{k} }  
                = t_1  \sum_{\mathbf{\delta}} \sin(\mathbf{k} \cdot \deltab ) ,
\label{eqBs}  \\  
  B^\sigma_{3,\mathbf{k} } &= \gamma_\sigma B_{3,\mathbf{k} } 
                = \gamma_\sigma(-2 t_2) \sin(\phi ) \sum_{\mathbf{\tau}} \sin(\mathbf{k} \cdot \taub ),
\nonumber	
\end{align}
with $\gamma_\uparrow = - \gamma_\downarrow = 1$ and 
the indices $\delta$ and $\tau$ corresponding to the
nearest-neighbor \eqref{deltavectors} and next-nearest-neighbor 
\eqref{tauvectors} vectors, respectively. 
Although the two matrices associated with each spin sector are different, 
they are not independent, since time-reversal symmetry
yields $h^\uparrow_\bk  =  h^{ \downarrow \, *}_{-\bk} $
(see Appendix A from Ref.~\cite{doretto2015flat} for further details).

It is possible to diagonalize the Hamiltonian \eqref{eqH0k} with the aid of
the canonical transformation
\begin{align}
 &c_{ \mathbf{k} A \uparrow} = u_{\mathbf{k}}^* d_{ \mathbf{k} \uparrow} +  v_{\mathbf{k}} c_{ \mathbf{k} \uparrow},   
\;\;
c_{ \mathbf{k}  A \downarrow} = u_{-\mathbf{k}} d_{ \mathbf{k} \downarrow} + v_{-\mathbf{k}}^* c_{ \mathbf{k} \downarrow}, 
\nonumber \\
 &c_{ \mathbf{k} B \uparrow}  = v_{\mathbf{k}}^{*} d_{\mathbf{k} \uparrow} - u_{\mathbf{k}} c_{ \mathbf{k} \uparrow}, 
\;\;
 c_{ \mathbf{k}  B \downarrow} = v_{-\mathbf{k}} d_{ \mathbf{k}\downarrow} - u_{-\mathbf{k}}^* c_{ \mathbf{k} \downarrow}, 
\label{eq:BogoTransf}
\end{align}
where the coefficients $u_\bk$ and $v_\bk$ are given by
\begin{align}	
 |u_{\mathbf{k}}|^2, |v_{\mathbf{k}}|^2 &= \frac{1}{2} \left(1 \pm \hat{B}_{3, \mathbf{k}} \right),  
\nonumber \\ 
 u_{\mathbf{k}} v_{\mathbf{k}}^{*} &=  \frac{1}{2} \left( \hat{B}_{1, \mathbf{k}} + i \hat{B}_{2, \mathbf{k}} \right),
\label{eq:Bogocoef} 
\end{align}
with $\hat{B}_{i,\bk}$ being the $i$-th component of the unit 
vector $\hat{\mathbf{B}}_\bk = \mathbf{B}_\bk/|\mathbf{B}_\bk|$. 
The diagonalized Hamiltonian reads  
\begin{align}
  H_{0} = \sum_{\mathbf{k} \sigma } 
              \omega_{\mathbf{k}}^c c_{\mathbf{k} \sigma}^{\dagger}  c_{\mathbf{k} \sigma}
           + \omega_{\mathbf{k}}^d d_{\mathbf{k} \sigma}^{\dagger}  d_{\mathbf{k} \sigma},
\label{eq:Hfree}
\end{align}
where the dispersions of the lower band $c$ ($-$ sign)  
and the upper band $d$ ($+$ sign) are given by
\begin{align}
    \omega^{d/c}_{\mathbf{k}} = B_0 \pm \sqrt{B_{1, \mathbf{k}}^2 + B_{2, \mathbf{k}}^2 + B_{3, \mathbf{k}}^2 } .
\label{eq:omega}
\end{align}
Notice that both $c$ and $d$ free-electronic bands are doubly
degenerated with respect to the spin degree of freedom.
For more details, 
we refer the reader to Fig.~2 from Ref.~\cite{leite2021}, where the  
free-electronic bands \eqref{eq:omega} are plotted  
for different values of the parameters $t_2/t_1$ and $\phi$.

As discussed in detail in Refs.~\cite{neupert2011fractional,leite2021}, 
the noninteracting band structure \eqref{eq:omega} have quite interesting
properties when the model parameters $t_2/t_1$ and $\phi$ are fine tunned. 
For instance, for (nearly flat band limit)
\begin{equation}
 \phi=0.656 \quad \text{and} \quad t_2 = 0.3155 t_1,
 \label{optimal-par}
\end{equation}
the lower band $c$ and the upper band $d$ are separated by an energy
gap and the lower band $c$ is almost flat. 
Away from the nearly flat band limit \eqref{optimal-par}, the lower
band $c$ gets more dispersive, and, in particular, for $\phi = 0$ or $t_2=0$, the
energy gap closes at the $K$ and $K'$ points of the first BZ   
(see Fig.~2(a) from Ref.~\cite{leite2021}).

In vicinity of the nearly flat band limit \eqref{optimal-par},
the free-electronic bands \eqref{eq:omega} are also topologically
nontrivial. Indeed, for tight-binding models of the form
\eqref{eqH0k}, one shows that the Chern numbers
of the upper and lower bands assume the form  
\cite{kane13,review-adv-phys18,rmp-class-top} 
\begin{equation}
 C_{\sigma}^{c/d} =\pm \gamma_\sigma\frac{1}{4 \pi} \int_{BZ} d^2k \, 
     \hat{\mathbf{B}}_\bk \cdot (\partial_{k_x} \hat{\mathbf{B}}_\bk \times \partial_{k_y} \hat{\mathbf{B}}_\bk ).  
\label{eqCn}
\end{equation}
In particular, for the noninteracting model \eqref{eqHH0}, 
one finds   
$C^d_\uparrow = -C^d_\downarrow = -1$ and 
$C^c_\uparrow = -C^c_\downarrow = +1$. 
As discussed in Sec.~IV from Ref.~\cite{doretto2015flat},
at half-filling, the so-called 
charge Chern number
$C_{\rm charge} = (C^c_\uparrow + C^c_\downarrow)/2 = 0$ 
while the spin Chern number 
$C_{\rm spin} = (C^c_\uparrow - C^c_\downarrow)/2 = 1$.
Since the tight-binding model \eqref{eqHH0} conserves the
$z$-component of the total spin (see Sec.~II.A from Ref.~\cite{doretto2015flat}),
the Z$_2$ topological invariants \cite{kane13, review-adv-phys18}
for the free-electronic bands 
$
  \nu_{c/d} = C^{c/d}_{\rm spin} \; {\rm mod} \; 2 = \pm 1,
$
i.e., the tight-binding model \eqref{eqHH0} is indeed a Z$_2$
topological insulator. At half filling, such a system should display
the quantum spin Hall  effect \cite{kane-mele05,kane2005quantum,zhang06} 
with the spin Hall conductivity  $\sigma^{SH}_{xy} = e C^c_{\rm spin}/2 \pi$.

%%%%%%%%%%%%%%%%%%%%%%%%%%%%%%%%%%%%%%%%%%%%%%%%%%%%%%%%%%%%%%%%%%%%%%%%%%%%%%%%%%%%%
\subsection{Interaction term in momentum space}
\label{sec:hubbard}

To find the expression of the on-site Hubbard repulsion term
\eqref{eqHHU} in momentum space, we start writing the Fourier
transform of the electron density operator \eqref{dens-op},
\begin{equation}
  \hat{\rho}_{i a \sigma} = \frac{1}{N} \sum_{\bq \in {\rm BZ}} 
                  e^{i \mathbf{q} \cdot \mathbf{R}_i} \hat{\rho}_{a \sigma}(\bq).
\label{eq:Fourier-rho}
\end{equation}
After substituting Eq.~\eqref{eq:Fourier-rho} into Eq.~\eqref{eqHHU},
we obtain  
\begin{equation}
   H_U = \frac{1}{N}\sum_{a = A,B} \sum_\bq U_a 
              \hat{\rho}_{a \uparrow}(-\bq) \hat{\rho}_{a \downarrow}(\bq).
\label{hu-k}
\end{equation} 

In terms of the fermion operators $c_{ \bk \, a \, \sigma}^{\dagger}$ 
[see Eq.~\eqref{eq:Fourier}],
the electron density operator $\hat{\rho}_{a \sigma}(\bq)$ reads 
\begin{equation}
   \hat{\rho}_{a \sigma}(\bq) = \sum_\bp c^\dagger_{\bp-\bq\, a\,\sigma}c_{\bp\, a\,\sigma}.
\label{density-op-2}
\end{equation}
Substituting Eq.~\eqref{eq:BogoTransf} into \eqref{density-op-2}
and neglecting the terms that contain the fermions 
$d_{\bk\,\sigma}$,  one finds the expression of the electron density
operator  \eqref{density-op-2} {\sl projected} into the lower noninteracting
bands $c$
(see Eq.~(28) from Ref.~\cite{doretto2015flat})
\begin{equation}
  \bar{\rho}_{a\, \sigma}(\bq) = \sum_\bp 
      G_{a\, \sigma}(\bp,\bq)c^\dagger_{\bp-\bq\,\sigma}c_{\bp\,\sigma},
\label{proj-dens-op} 
\end{equation}
where the $G_{a\,\sigma}(\bp,\bq)$ functions are given by 
\begin{align}
 G_{a\, \sigma}(\bp, \bq) &= 
          \delta_{a,A} \left( \delta_{\sigma,\uparrow}v^*_{\bp - \bq} v_\bp 
                                 + \delta_{\sigma,\downarrow}v_{-\bp + \bq} v^*_{-\bp} \right)
\nonumber \\
    &+ \delta_{a,B} \left( \delta_{\sigma,\uparrow}u^*_{\bp - \bq} u_\bp 
                  + \delta_{\sigma,\downarrow}u_{-\bp + \bq} u^*_{-\bp} \right),
\label{eq:ga-sigma}
\end{align}
with $u_\bk$ and $v_\bk$ being the coefficients \eqref{eq:Bogocoef}. 

Finally, we quote the expression of the on-site Hubbard
term \eqref{hu-k} projected into the lower noninteracting bands $c$, which
follows from Eq.~\eqref{hu-k} 
with $\hat{\rho}_{a \sigma}(\bq) \rightarrow \bar{\rho}_{a \sigma}(\bq)$: 
\begin{equation}
   \bar{H}_U = \frac{1}{N}\sum_{a = A,B} \sum_\bq U_a 
              \bar{\rho}_{a \uparrow}(-\bq) \bar{\rho}_{a \downarrow}(\bq).
\label{hu-k-bar}
\end{equation}

%%%%%%%%%%%%%%%%%%%%%%%%%%%%%%%%%%%%%%%%%%%%%%%%%%%%%%%%%%%%%%%%%%%%%%%%%%%%%%%%%%%%% 
\section{Bosonization formalism for flat-band Z$_2$ topological insulators}
\label{sec:boso}

Here we summarize the bosonization formalism for a Z$_2$
topological insulator introduced by one of us in Ref.~\cite{doretto2015flat}
for the description of the flat-band FM phase of a
square lattice correlated Z$_2$ topological insulator.  
Our discussion follows the lines of Sec.~III from Ref.~\cite{leite2021}.

In order to introduce the bosonization scheme, one needs to define 
a reference state.
Let us consider a spinfull topological insulator on a bipartite lattice whose
Hamiltonian assumes the form \eqref{eqH0k}, choose the model
parameters such that (at least) the lower band $c$ is (nearly)
flat, and focus on the $1/4$ filling of the electronic bands:
the number of electrons 
$N_e = N_A = N_B = N$, with $N_A$ and $N_B$ being, respectively, the
number of sites of the sublattices $A$ and $B$. 
Assuming that the lower band $c \, \uparrow\,$ 
is completely occupied [see Fig.~\ref{fig:Lattice}(c)],    
the ground state of the noninteracting system 
(the {\sl reference} state) is completely spin polarized:
\begin{equation}
 |{\rm FM} \rangle =  \prod_{\mathbf{k} \in BZ} c_{\mathbf{k} \uparrow}^{\dagger} |0 \rangle.
\label{eq:FM}
\end{equation}
Excited states are generated by spin-flips: As illustrated in
Fig.~\ref{fig:Lattice}(c), the lowest-energy neutral excitations
above the reference state \eqref{eq:FM} are particle-hole pairs
within the lower bands $c$, since the lower flat bands $c$ are
separated from the upper ones $d$ by an energy gap; such an excited state
with well-defined momentum can be written as
$| \Psi_{\mathbf{k}} \rangle \propto S_\bk^{-} | {\rm FM} \rangle$.
Interestingly, it is possible to define boson operators that are
associated with such spin-flip excitations 
(see Ref.~\cite{doretto2015flat} for details), 
\begin{align}
  b_{\alpha,\mathbf{q}} &= \frac{\bar{S}_{-\mathbf{q},\alpha}^{+}}{F_{\alpha\alpha,\mathbf{q}}} 
                = \frac{1}{F_{\alpha\alpha,\bq}}  \sum_{\mathbf{p}} g_{\alpha}^* (-\mathbf{p}, \mathbf{q}) 
                   c_{\mathbf{p+q}\uparrow}^{\dagger} c_{\mathbf{p}\downarrow},
\nonumber \\
   b_{\alpha,\mathbf{q}}^{\dagger} &= \frac{\bar{S}_{\mathbf{q},\alpha}^{-}}{F_{\alpha\alpha,\mathbf{q}}} 
             =  \frac{1}{F_{\alpha\alpha,\bq} }  \sum_{\mathbf{p}} g_{\alpha} (\mathbf{p}, \mathbf{q}) 
                 c_{\mathbf{p-q}\downarrow}^{\dagger} c_{\mathbf{p}\uparrow}, 
\label{eq:bosons}
\end{align}
with $\alpha = 0,1$, that obey the commutation relations
\begin{align}
 [b_{\alpha,\mathbf{k}} , b_{\beta, \mathbf{q}}^{\dagger}  ] &= \delta_{\alpha, \beta} \; \delta_{\mathbf{k}, \mathbf{q}}, 
\nonumber  \\ 
 [b_{\alpha,\mathbf{k}} , b_{\beta,\mathbf{q}}  ] &=  [b_{\alpha,\mathbf{k}}^{\dagger} , b_{\beta,\mathbf{q}}^{\dagger}  ] = 0.
\label{eq:BComutations}
\end{align}

Concerning the definition of the projected spin operators $\bar{S}^\pm_{\mathbf{q}, \alpha}$
in Eq.~\eqref{eq:bosons}, we consider {\sl two distinct} proposals: \\
(i) {\bf Mixed-lattice excitations}: Motivated by previous results \cite{doretto2015flat} 
concerning a correlated Z$_2$ topological insulator on a square lattice,
we define $\bar{S}^\pm_{\mathbf{q}, \alpha}$ as 
\begin{equation}
\bar{S}^\pm_{\bq, \alpha}  = \bar{S}^\pm_{\bq, AB}
                    + (-1)^\alpha \bar{S}^\pm_{\bq, BA},
\label{eq:projS1}
\end{equation}
where 
\begin{equation}
  \bar{S}^\pm_{\bq,ab} = \bar{S}^x_{\bq,ab} \pm i \bar{S}^y_{\bq,ab}.
\label{linearcomb}
\end{equation}
The operator $\bar{S}^\lambda_{\bq,ab}$,
with $\lambda = x,y,z$ and $a,b = A$, $B$,  
is the spin operator $S^\lambda_{\bq,ab}$ {\sl projected} 
into the lower noninteracting bands $c$.
The spin operator $S^\lambda_{\bq,ab}$ is indeed the Fourier transform of
the operator
\begin{equation}
   S^\lambda_{i, ab} = \frac{1}{2}\sum_{\mu,\nu=\uparrow,\downarrow} 
                              c_{i a \mu}^\dagger \sigma^\lambda_{\mu\,\nu} c_{i b \nu},
\label{eq:abrikosov}
\end{equation}
where $\sigma^\lambda_{\mu\,\nu} $ is the matrix element of the Pauli
matrix $\sigma^\lambda$.  
The spin operators \eqref{eq:projS1} are indeed related
with spin-flip excitations that also change the sublattice index.
Due to such a feature, 
we denote the excitations defined by  
the boson operators \eqref{eq:bosons} and the spin operator \eqref{eq:projS1}
as {\sl mixed-lattice} (ML) excitations.
The $F^2_{\alpha\beta, \bq}$ function reads  
\begin{align}
   F^2_{\alpha \beta, \bq} = \sum_\bp g_\alpha(\bp, \bq) g_\beta^*(-\bp+\bq,\bq),
\label{eq:F2abhigh}
\end{align} 
with $g_\alpha(\mathbf{p},\mathbf{q})$ defined in terms
of the coefficients \eqref{eq:Bogocoef},
\begin{align}
   g_\alpha(\bp,\bq) =  -u_\bp v_{-\bp+\bq} - (-1)^\alpha v_\bp u_{-\bp + \bq}.
\label{eq:g1}
\end{align}
Interestingly, the $F^2_{\alpha \beta, \bq}$ function can be explicitly expressed 
in terms of the $B_{i, \mathbf{k}}$ functions \eqref{eqBs}, see Eq.~\eqref{eq:F2high}. \\
(ii) {\bf Same-lattice excitations}: Motivated by our previous study \cite{leite2021}    
about a honeycomb lattice correlated Chern insulator,   
we also consider spin-flip excitations that preserve the sublattice
index. In this case, one defines 
\begin{equation}
   \bar{S}^\pm_{\mathbf{q}, \alpha} = \bar{S}^\pm_{\mathbf{q}, A}+ (-1)^\alpha \bar{S}^\pm_{\mathbf{q}, B},
\label{eq:projS2}
\end{equation}
where $\bar{S}^\pm_{\bq, a}$ is also 
given by Eqs.~\eqref{linearcomb} and \eqref{eq:abrikosov}
with $a = b$, i.e., $\bar{S}^\pm_{\bq, a} = \bar{S}^\pm_{\bq, aa}$;
again, boson operators are defined as done in Eq.~\eqref{eq:bosons},
with $F^2_{\alpha \beta, \bq}$ also given by Eq.~\eqref{eq:F2abhigh},
but now $g_\alpha(\mathbf{p},\mathbf{q})$ assumes 
the form
\begin{align}
  g_\alpha(\bp,\bq) =  v_{-\bp+\bq} v_\bp + (-1)^\alpha u_{-\bp + \bq} u_\bp,
\label{eq:g2}
\end{align}
with $u_\bk$ and $v_\bk$ being the coefficients \eqref{eq:Bogocoef}. 
Since the spin operators \eqref{eq:projS2} preserve the sublattice
index, we denote such excitations as  {\sl same-lattice} (SL) excitations.
The expression of the $F^2_{\alpha \beta, \bq}$ function in terms of the 
$B_{i, \mathbf{k}}$ functions \eqref{eqBs} is shown in 
Appendix~\ref{sec:ap-details-same},
see Eq.~\eqref{eq:F2low}.  
Finally, one should note that, for both ML and SL excitations,
\begin{equation}
   b_{\alpha,\bq} | {\rm FM} \rangle = 0,
\label{vacuum} 
\end{equation}
which indicates that the spin-polarized (reference) state
\eqref{eq:FM} is indeed the vacuum for the boson operators
\eqref{eq:bosons}, regardless the definition of the projected spin operators.

As discussed in detail in Ref.~\cite{doretto2015flat}, 
it is possible to find the bosonic representation of
any operator that is written in terms of the fermions 
$c^\dagger_{\bk\sigma}$ and $c_{\bk\sigma}$.
For instance, in terms of the boson operators \eqref{eq:bosons}
(either defined in terms of the ML or the SL excitations), 
the bosonic representation of the projected electron density operator
\eqref{proj-dens-op} reads 
\begin{equation}
  \bar{\rho}_{a \sigma}(\mathbf{k}) = \frac{1}{2}N\delta_{\sigma, \uparrow}\delta_{\mathbf{k}, 0} 
        + \sum_{\alpha,\beta,\bq} \: \mathcal{G}_{\alpha \beta a \sigma}(\mathbf{k}, \mathbf{q})  
        b_{\beta,\mathbf{k}+\mathbf{q} }^{\dagger} b_{\alpha, \mathbf{q}}, 
\label{eq:rhoBoson}
\end{equation}
where the $\mathcal{G}_{\alpha \beta a \sigma}(\bk,\bq)$ function is 
defined by Eq.~\eqref{Gcal}. 
Similar to the  $F^2_{\alpha \beta, q}$ function \eqref{eq:F2abhigh}, 
$\mathcal{G}_{\alpha \beta a \sigma}(\bk,\bq)$ 
can also be written in terms of the coefficients \eqref{eqBs}, 
see Eqs.~\eqref{Gcal2}  and \eqref{Gcal3} for ML and SL
excitations, respectively.  
As discussed in the next section, both the   
Hamiltonian \eqref{eqHH0} and the interaction term
\eqref{eqHHU}, projected into the lower noninteracting bands $c$,
can also be expressed in terms of the boson operators \eqref{eq:bosons}. 
Apart from the expressions of $F^2_{\alpha \beta, q}$ and 
$\mathcal{G}_{\alpha\beta a \sigma}(\bk,\bq)$,    
the bosonic representation \eqref{eq:rhoBoson} of the density operator
\eqref{proj-dens-op} and the effective boson model [see Eq.~\eqref{eq:Heffective} below] 
derived from the THM \eqref{eqHH} are equal,
regardless the nature of the excitations considered (ML or SL ones); 
due to such a feature, we employ the same notation for the boson
operators \eqref{eq:bosons} for both ML \eqref{eq:projS1}
and SL \eqref{eq:projS2} excitations.

Finally, it is important to emphasize that, for the square lattice $\pi$-flux model
\cite{doretto2015flat}, only the ML excitations \eqref{eq:projS1} 
yield two sets of independent bosons operators $b_0$ and $b_1$. 
Such a feature distinguishes the time-reversal symmetric square lattice
$\pi$-flux model from the generalized Haldane one [Eq.~\eqref{eqHH0}],
which, in principle, allows us to define boson operators from both 
ML \eqref{eq:projS1} and SL \eqref{eq:projS2} excitations. 
Interestingly, for the generalized square lattice $\pi$-flux model
\cite{doretto2015flat} and the generalized Haldane model \cite{leite2021}, 
both with broken time-reversal symmetry,  
the SL excitations \eqref{eq:projS2} are the
lowest-energy excitations of the corresponding correlated Chern
insulators.

%%%%%%%%%%%%%%%%%%%%%%%%%%%%%%%%%%%%%%%%%%%%%%%%%%%%%%%%%%%%%%%%%%%%%%%%%%%%%%%%%%%%%	
\section{Flat-band ferromagnetism in the topological Hubbard model}
\label{sec:flatferromagnetism}

In this section, we study the flat-band FM phase of the THM \eqref{eqHH}.   
We consider the model at $1/4$ filling of its corresponding
noninteracting limit and assume that the noninteracting lower bands
$c$ are in the vicinity of the nearly flat band limit \eqref{optimal-par}.
We focus on the determination of the dispersion
relation of the elementary particle-hole pair excitations 
(spin-waves) above the (flat-band) FM ground state \eqref{eq:FM}: 
ML [Eq.~\eqref{eq:projS1}] and SL [Eq.~\eqref{eq:projS2}] 
excitations are discussed separately, since they are two distinct
proposals for the definition of the boson operators \eqref{eq:bosons};
most importantly, we find that the SL excitations \eqref{eq:projS2} are
indeed the lowest-energy excitations.

%%%%%%%%%%%%%%%%%%%%%%%%%%%%%%%%%%%%%%%%%%%%%%%%%%%%%%%%%%%%%%%%%%%%%%%%%%%%%%%%%%%%% 
\subsection{Effective interacting boson model}
\label{sec:ChernInsu}

Here we derive an effective interacting boson model from
the THM \eqref{eqHH} 
within the bosonization formalism summarized in Sec.~\ref{sec:boso}. 
Our presentation closely 
follows the lines of Sec.~IV.A from Ref.~\cite{leite2021} 
and more details can be found in Ref.~\cite{doretto2015flat}.

First of all, we project the Hamiltonian \eqref{eqHH} into the
lower noninteracting bands $c$ 
(such a restriction is justified, once the on-site repulsion energies
$U_a$ fullfil some conditions, see comment above Eq.~(35) from
Ref.~\cite{leite2021}),  
\begin{align}
      H \rightarrow \bar{H} &= \bar{H}_{0} + \bar{H}_U,
\label{H-projected}
\end{align}	
where the projected noninteracting Hamiltonian $\bar{H}_{0}$ is
obtained from Eq.~\eqref{eq:Hfree}, 
\begin{align}
  \bar{H}_{0} = \sum_{\mathbf{k} \sigma }  \omega^c_{ \mathbf{k}}
                     c_{\mathbf{k} \sigma}^{\dagger}  c_{\mathbf{k} \sigma},	
\label{eq:omegaProje}
\end{align}
and $\bar{H}_U$ is given  Eq.~\eqref{hu-k-bar}. 
In terms of the boson operators \eqref{eq:bosons}, 
the noninteracting (kinetic) term $\bar{H}_{0}$ reads
\begin{equation}
    \bar{H}_{0,B} = E_0 + \sum_{\alpha \beta} \sum_{\mathbf{q} \in BZ}  \bar{\omega}^{\alpha \beta}_{\mathbf{q} }
                           b_{\beta, \mathbf{q}}^{\dagger} b_{\alpha, \mathbf{q}},
\label{eq:H0B1}
\end{equation}
where $E_0 = \sum_\bk \omega^c_\bk$ is a constant related to  
the action of the Hamiltonian $\bar{H}_0$ into the reference state
\eqref{eq:FM} and 
\begin{align}
    \bar{\omega}^{\alpha \beta}_{\mathbf{q}}  &=  
            \sum_{\mathbf{p}}
            \left( \omega^c_{\mathbf{p-q}} - \omega^c_{\mathbf{p}} \right) 
            \frac{g_{\alpha} (\mathbf{p}, \mathbf{q})  g^*_{\beta}(-\bp+\bq, \bq)}
                    {F_{\alpha\alpha, \mathbf{q}} F_{\beta\beta, \mathbf{q}}}, 
\label{eq:omegaBar} 
\end{align}
with $F_{\alpha\beta, \bq}$ given by 
Eqs.~\eqref{eq:F2high} (ML excitations) and
\eqref{eq:F2low} (SL excitations) and  $g_{\alpha}(\mathbf{p}, \mathbf{q})$ 
given by Eqs.~\eqref{eq:g1} (ML excitations) and \eqref{eq:g2} (SL excitations).
The on-site Hubbard term $\bar{H}_U$ can be cast into its bosonic
representation with the aid of Eqs.~\eqref{hu-k-bar} and \eqref{eq:rhoBoson}; 
after normal-ordering the resulting expression, 
one arrives at \cite{doretto2015flat} 
\begin{align}
     \bar{H}_{U,B}  &= \bar{H}_{U,B}^{(2)}+ \bar{H}_{U,B}^{(4)},
\end{align}
where the quadratic and quartic terms are given by
\begin{align}
 \bar{H}_{U,B}^{(2)} &=  \sum_{\alpha \beta} \sum_{\mathbf{q} } \epsilon^{\alpha \beta}_{\mathbf{q} }
                                   b_{\beta, \mathbf{q}}^{\dagger} b_{\alpha, \mathbf{q}},
\label{H42} \\
 \bar{H}_{U,B}^{(4)} &= \frac{1}{N} \sum_{\mathbf{k} , \mathbf{q}, \mathbf{p}} \sum_{\alpha \beta \alpha' \beta'} 
                V^{\alpha \beta \alpha' \beta'}_{\mathbf{k}, \mathbf{q}, \mathbf{p} }
                        b_{\beta', \mathbf{p+k}}^{\dagger} b_{\beta, \mathbf{q-k}}^{\dagger} b_{\alpha \mathbf{q}} b_{\alpha' \mathbf{p}}, 
\label{H44}
\end{align}
with the coefficient $\epsilon^{\alpha \beta}_\bq $ assuming the form
\begin{align}
 \epsilon^{\alpha \beta}_\bq &=  \frac{1}{2} \sum_{a}
                          U_a\mathcal{G}_{\alpha \beta a \downarrow} (0, \mathbf{q}) 
\nonumber \\
                       &+ \frac{1}{N} \sum_{a,\alpha', \mathbf{k}}
                         U_a\mathcal{G}_{\alpha' \beta a \uparrow}(-\mathbf{k}, \mathbf{k+q}) 
                           \mathcal{G}_{\alpha \alpha' a \downarrow}(\mathbf{k}, \mathbf{q}),
\label{eq:Epsilon}
\end{align}
and the boson-boson interaction being defined by
\begin{align}
         V^{\alpha \beta \alpha' \beta'}_{\mathbf{k}, \mathbf{q}, \mathbf{p} } &= \frac{1}{N} \sum_a 
                       U_a\mathcal{G}_{\alpha \beta a \uparrow}(-\mathbf{k}, \mathbf{q}) 
                       \mathcal{G}_{\alpha' \beta' a \downarrow} (\mathbf{k}, \mathbf{p}). 
\label{eq:Vkq}
\end{align}
One should recall that, in terms of the coefficients \eqref{eqBs}, the 
$\mathcal{G}_{\alpha \beta a \sigma}(\bk,\bq)$ functions are given by
Eqs.~\eqref{Gcal2}  and \eqref{Gcal3} for ML and SL
excitations, respectively. 
In summary, the effective {\sl interacting} boson model,     
which allows us to describe the flat-band FM phase of the
THM \eqref{eqHH}, reads 
\begin{align}
  \bar{H}_B = \bar{H}_{0,B} + \bar{H}^{(2)}_{U,B} + \bar{H}^{(4)}_{U,B}.
\label{eq:Heffective}
\end{align}

It is important to emphasize that the effective boson model
\eqref{eq:Heffective} is quite general, since, in principle, 
it can describe the flat-band FM phase of a correlated Z$_2$ 
topological insulator described by a THM  
on a bipartite lattice, as long as its corresponding noninteracting term assumes
the form \eqref{eqH0k} and its free-electronic bands can be
made almost dispersionless by carefully choosing the model
parameters (see Sec.~V from Ref.~\cite{leite2021} for more details):
recall that, all terms of the Hamiltonian \eqref{eq:Heffective}
can be written in terms of the  functions \eqref{eqBs}, which
completely characterize tight-binding models of the form \eqref{eqH0k}.

%%%%%%%%%%%%%%%%%%%%%%%%%%%%%%%%%%%%%%%%%%%%%%%%%%%%%%%%%%%%%%%%%%%%%%%%%%%%%%%%%%%%%
\subsection{Spin-wave spectrum}
\label{sec:spin-wave}

We now determine the spin-wave spectrum of the flat-band
FM phase of the THM \eqref{eqHH} with the aid of 
effective boson model \eqref{eq:Heffective}. In the
lowest-order (harmonic) approximation, 
the Hamiltonian \eqref{eq:Heffective} reads
\begin{align}
  \bar{H}_B \approx \bar{H}_{0,B} + \bar{H}^{(2)}_{U,B}.
\label{eq:Heffective2}
\end{align}

The Hamiltonian \eqref{eq:Heffective2} can be diagonalized with the
aid of the following canonical transformation
\begin{equation}
   b_{0, \bq } = u^*_\bq a_{+, \bq} + v_\bq a_{-, \bq},
\quad
   b_{1, \bq } = v^*_\bq a_{+, \bq} - u_\bq a_{-, \bq}.
\label{eq:BogoTransf2}
\end{equation}
One then easily shows that 
\begin{equation}
  \bar{H}_B  =  E_0 +   \sum_{\mu = \pm } \sum_{\bq \in BZ }   
   \Omega_{\mu, \bq} a_{\mu, \mathbf{q}}^{\dagger} a_{\mu, \mathbf{q}},
\label{eq:HFinalFlat}
\end{equation}
where the constant 
$E_0 = \sum_\bk \omega^c_\bk = (-1.69\,t_1)N$ for the nearly flat band
limit \eqref{optimal-par}, the dispersion relation $\Omega_{\mu, \bq}$ 
of the bosons $a_\pm$ (the spin-wave spectrum) is given by
\begin{equation}
  \Omega_{\pm,\bq} = \frac{1}{2}\left( \epsilon^{00}_\bq + \epsilon^{11}_\bq \right) 
                                 \pm \epsilon_\bq,
\label{omega-b}
\end{equation}
with 
$\epsilon_\bq = \frac{1}{2}\sqrt{ \left( \epsilon^{00}_\bq - \epsilon^{11}_\bq \right)^2
                       +  4 \epsilon^{01}_\bq \epsilon^{10}_\bq}$,
and the coefficients $u_\bq$ and $v_\bq$ satisfy the relations 
\begin{align}	
 |u_\bq|^2,  |v_\bq|^2  &=   \frac{1}{2}  \pm
          \frac{1}{4\epsilon_\bq}\left( \epsilon^{00}_\bq - \epsilon^{11}_\bq \right),
\nonumber \\
 u_\bq v_\bq^{*} &=   \frac{\epsilon^{01}_\bq}{4\epsilon_\bq},
\quad
 v_\bq u_\bq^{*}  =   \frac{\epsilon^{10}_\bq}{4\epsilon_\bq}.
\label{eq:Bogocoef2} 
\end{align}
Note that the vacuum state for the bosons $a_\pm$ is the ground state of the
Hamiltonian \eqref{eq:HFinalFlat}. Indeed, due to the form of the canonical 
transformation \eqref{eq:BogoTransf2}, one sees that the vacuum for
the bosons $a_\pm$ corresponds to the spin-polarized ferromagnet state
\eqref{eq:FM}, which is the vacuum
(reference) state for the bosons $b_0$ and $b_1$ [see Eq.~\eqref{vacuum}].
Such a result points to the stability of a flat-band FM
phase for the THM \eqref{eqHH}. 

The spin-wave spectra \eqref{omega-b} for the ML excitations
\eqref{eq:projS1} are shown in Figs.~\ref{figEspectro}(a)--(c),  
while the results for the SL ones \eqref{eq:projS2} are displayed in 
Figs.~\ref{figEspectro}(d)--(f) and \ref{fig:espectro-phi}(a) and (b). 
In the following, we concentrate on the spin-wave spectrum for the SL
excitations, since they are the lowest-energy excitations
that characterize the flat-band FM phase of the THM 
\eqref{eqHH}. A detailed discussion about the ML excitations
can be found in Appendix~\ref{sec:ap-spin-wave}.

%%%%%%%%%%%%%%%%%%%%%%%%%%%%%%%%%%%%%%%%%%%%%%%%%%%%%%%%%%%%%%%%%%%%%%%%%%%%%%%%%%%%% 
\subsubsection*{SL excitations}
\label{sec:spin-waveII}

In order to determine the spin-wave spectrum
\eqref{omega-b} for the SL excitations \eqref{eq:projS2},
one needs to calculate the kinetic coefficients \eqref{eq:omegaBar} 
and the coefficients \eqref{eq:Epsilon}
associated with the quadratic term \eqref{H42}.
In this case, one should consider the expressions of the
$g_\alpha(\bp, \bq)$, $F_{\alpha\beta, \bq}$, and 
$\mathcal{G}_{\alpha \beta a \sigma}(\bp,\bq)$ 
functions given by Eqs.~\eqref{eq:g2},
\eqref{eq:F2low}, and \eqref{Gcal3}, respectively. 
Differently from the ML excitations (see Appendix~\ref{sec:ap-spin-wave}), 
for the SL excitations, one finds that the kinetic coefficients
\eqref{eq:omegaBar} vanishes,  $\bar{\omega}^{\alpha \beta}_{\mathbf{q}} = 0$.
Moreover, the quadratic term \eqref{H42} of the effective boson model 
\eqref{eq:Heffective} is Hermitian, since the coefficients
$\epsilon^{\alpha \alpha}_\bq $ are real quantities 
while  $\epsilon^{01}_\bq$ and $\epsilon^{10}_\bq$ are complex
ones with $\epsilon^{01}_\bq = (\epsilon^{10}_\bq)^*$ 
[see Eq.~\eqref{eq:Epsilon} and Fig.~\ref{fig:F2low}(c)];
such a feature is distinct from the ones found for the ML excitations 
(see Appendix~\ref{sec:ap-spin-wave}) and for the correlated
Chern insulator \cite{leite2021}, whose corresponding 
quadratic Hamiltonians \eqref{H42} are non-Hermitian.
Finally, similarly to the ML excitations (see Fig.~\ref{fig:F2high}) 
and the correlated Chern insulator \cite{leite2021},
one finds that the condition
\begin{equation}
  F_{\alpha \beta,\bq} = \delta_{\alpha,\beta}F_{\alpha \alpha,\bq}
\label{conditionF}
\end{equation}
is not fulfilled for all momenta within the first BZ 
[see Figs.~\ref{fig:F2low} (a) and (b)]; 
the validity of the condition \eqref{conditionF} is an important
ingredient for the definition \eqref{eq:bosons} 
of the two sets of {\sl independent} boson operators $b_0$ and $b_1$;
for a detailed discussion about this important issue, we refer the
reader to Appendix~\ref{sec:ap-spin-wave}
and to Appendix~B from Ref.~\cite{leite2021}.

\begin{figure*}[t]
\centerline{\includegraphics[width=6.1cm]{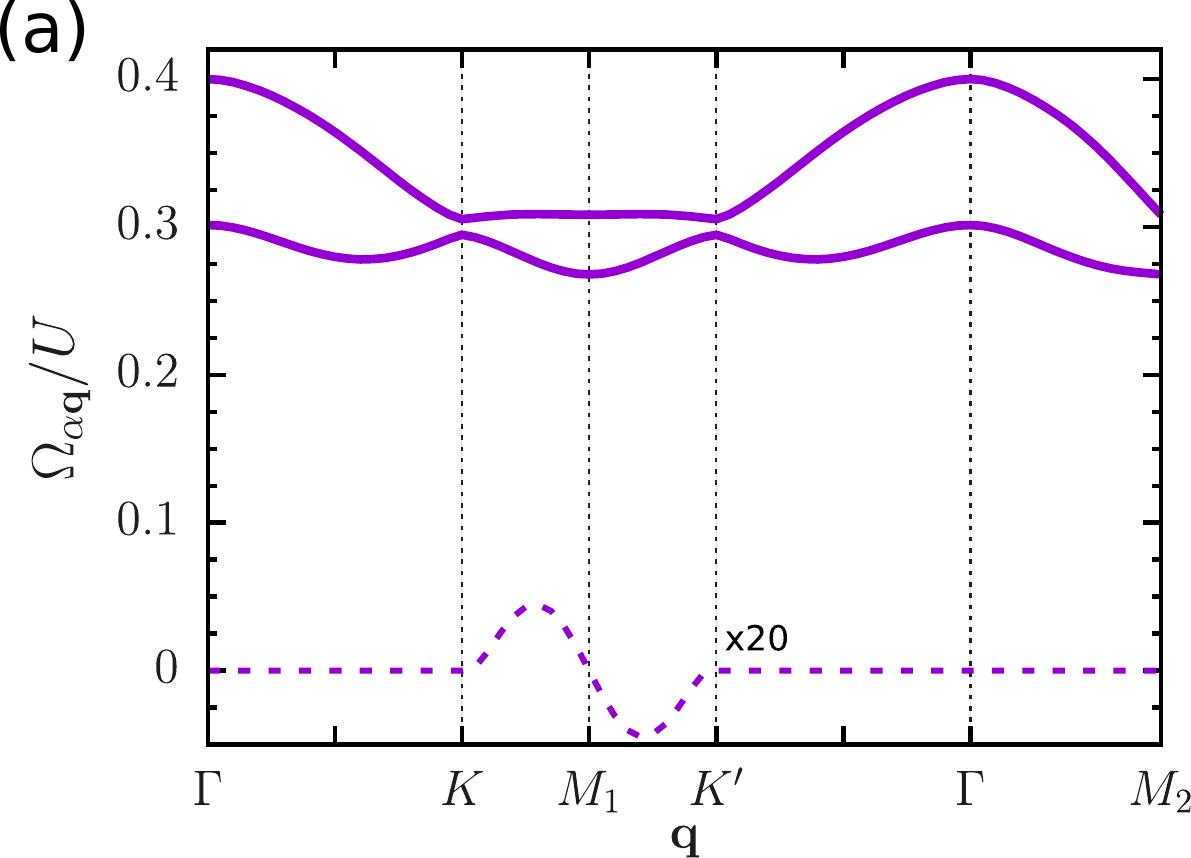}
                   \hskip2.0cm \includegraphics[width=6.1cm]{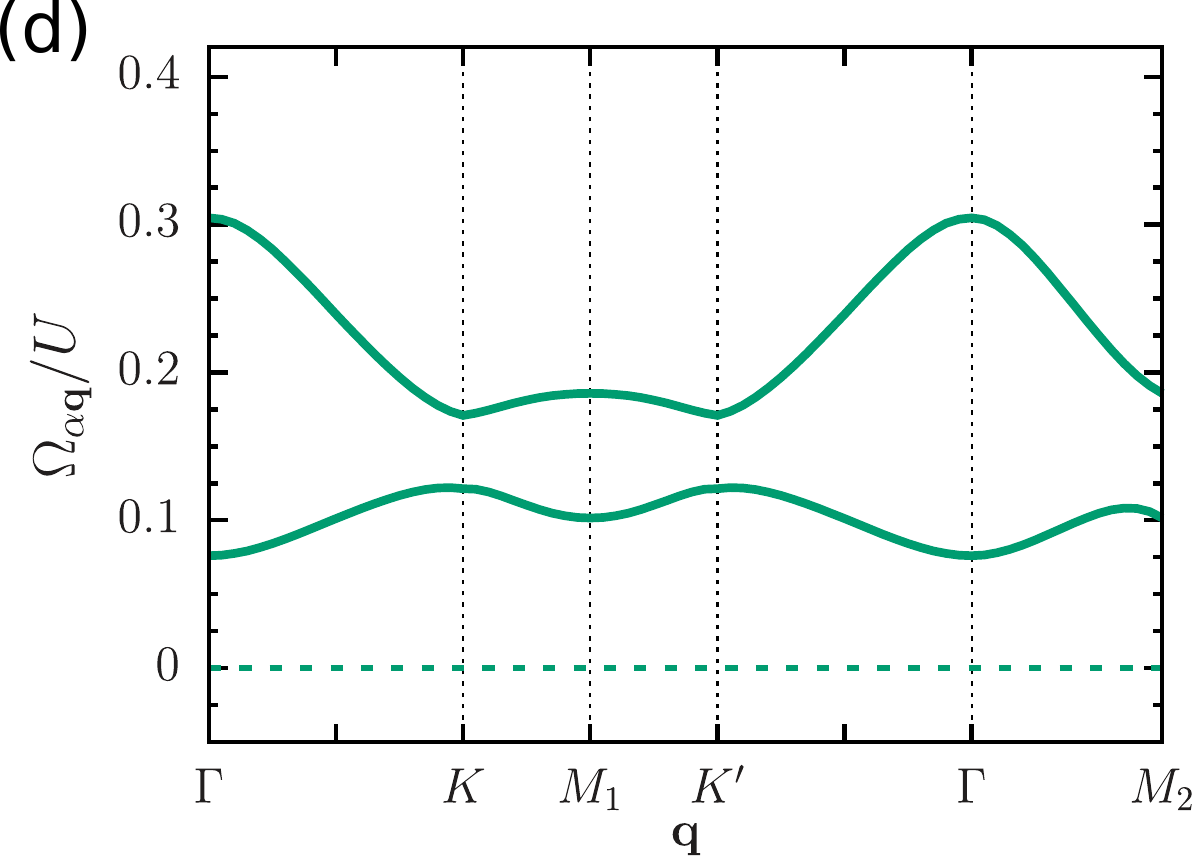}}
\vskip0.2cm
\centerline{\includegraphics[width=6.1cm]{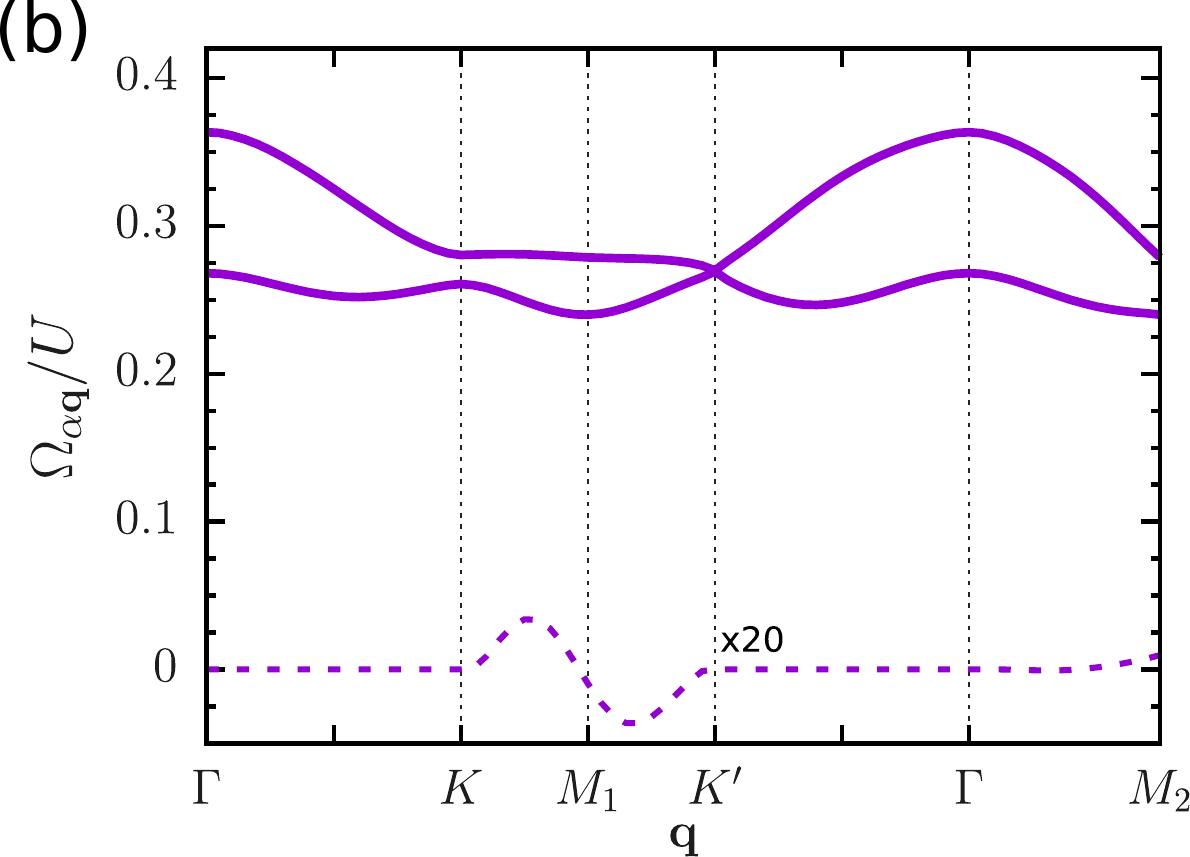}
                   \hskip2.0cm \includegraphics[width=6.1cm]{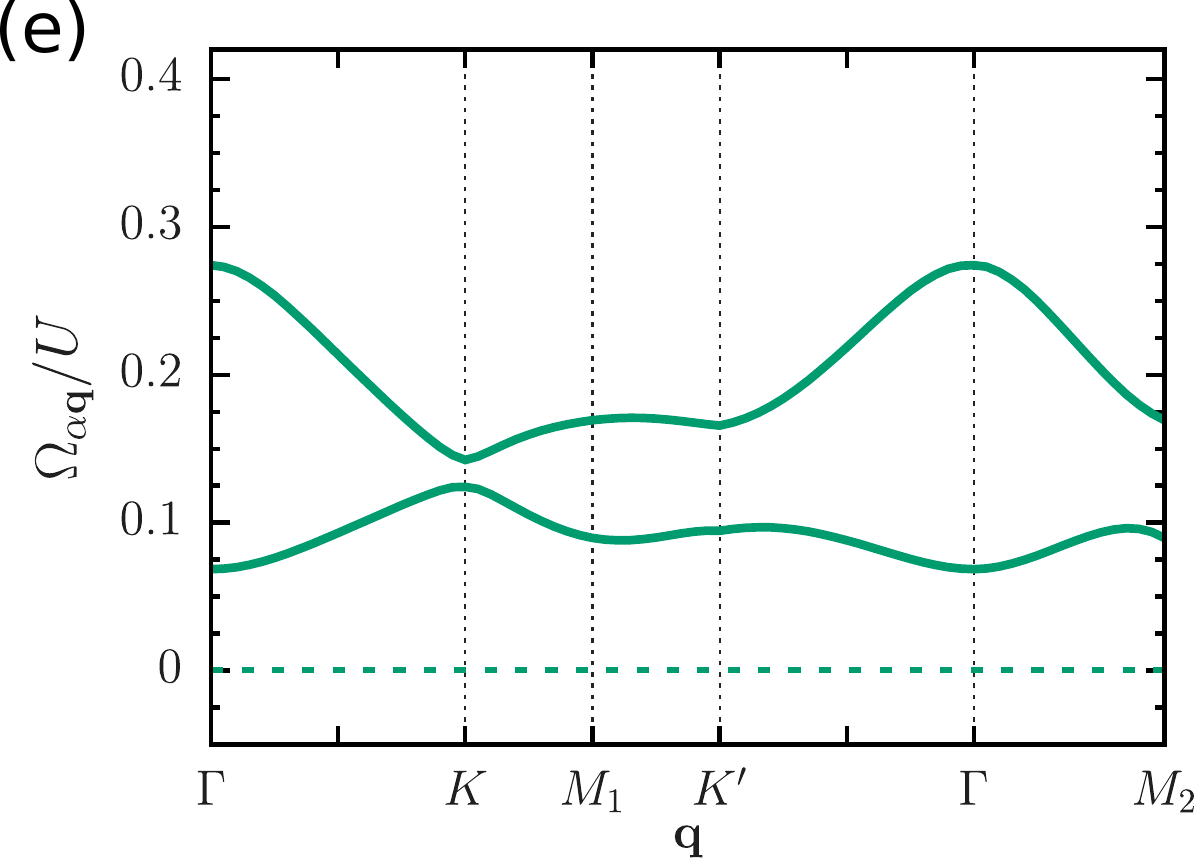}}
\vskip0.2cm
\centerline{\includegraphics[width=6.1cm]{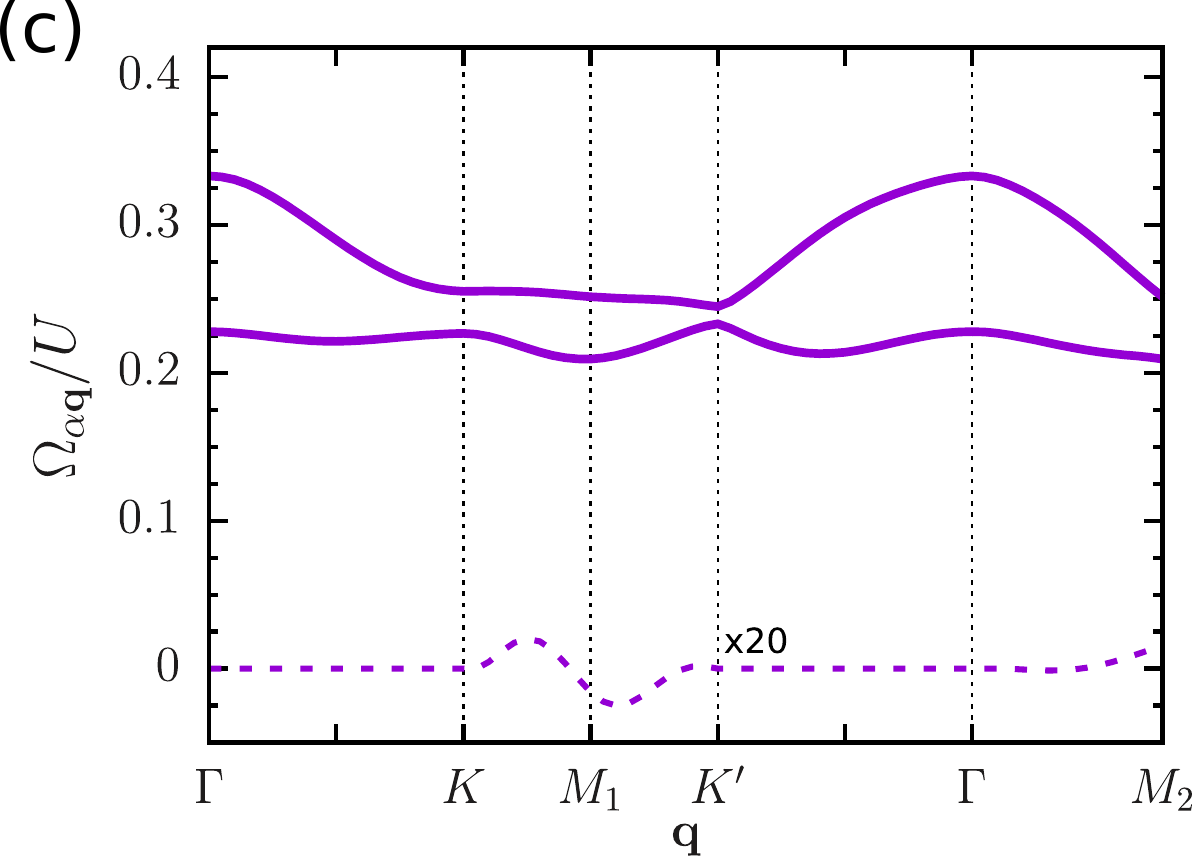}
                   \hskip2.0cm \includegraphics[width=6.1cm]{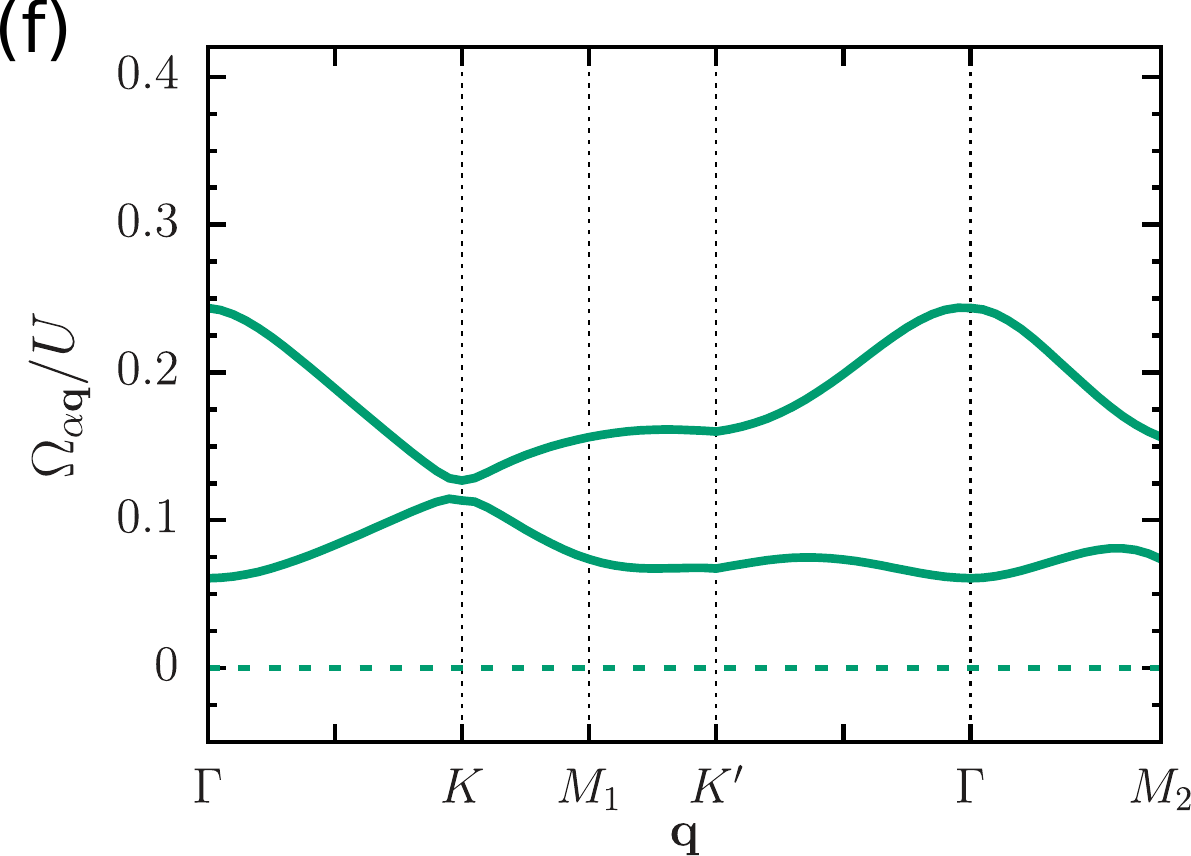}
}
\caption{Dispersion relation \eqref{omega-b} (spin-wave spectrum) of
the effective boson model \eqref{eq:Heffective2} in the harmonic
approximation for the nearly flat band limit \eqref{optimal-par}
along paths in the first BZ  [Fig.~\ref{fig:Lattice}(b)].
Solid and dashed lines respectively represent the real part of $\Omega_{\pm,\bq}$
and the imaginary part of $\Omega_{+,\bq} = -\Omega_{-,\bq} $, where
the latter is multiplied by a factor of 20 for clarity.  
The spin-wave spectrum (solid magenta line) for the ML excitations \eqref{eq:projS1} are
shown in panels (a), (b), and (c), while 
panels (d), (e), and (f) correspond to the spin-wave spectrum (solid
green line) for the SL excitations \eqref{eq:projS2}.
The on-site Hubbard repulsion energies are  
$U_A = U_B = U$ [(a) and (d)],
$U_B = 0.8\, U_A = 0.8\, U$ [(b) and (e)], and 
$U_B = 0.6\, U_A = 0.6\, U$ [(c) and (f)]. } 
\label{figEspectro}
\end{figure*}

\begin{figure}[t] 
\centerline{\includegraphics[width=5.5cm]{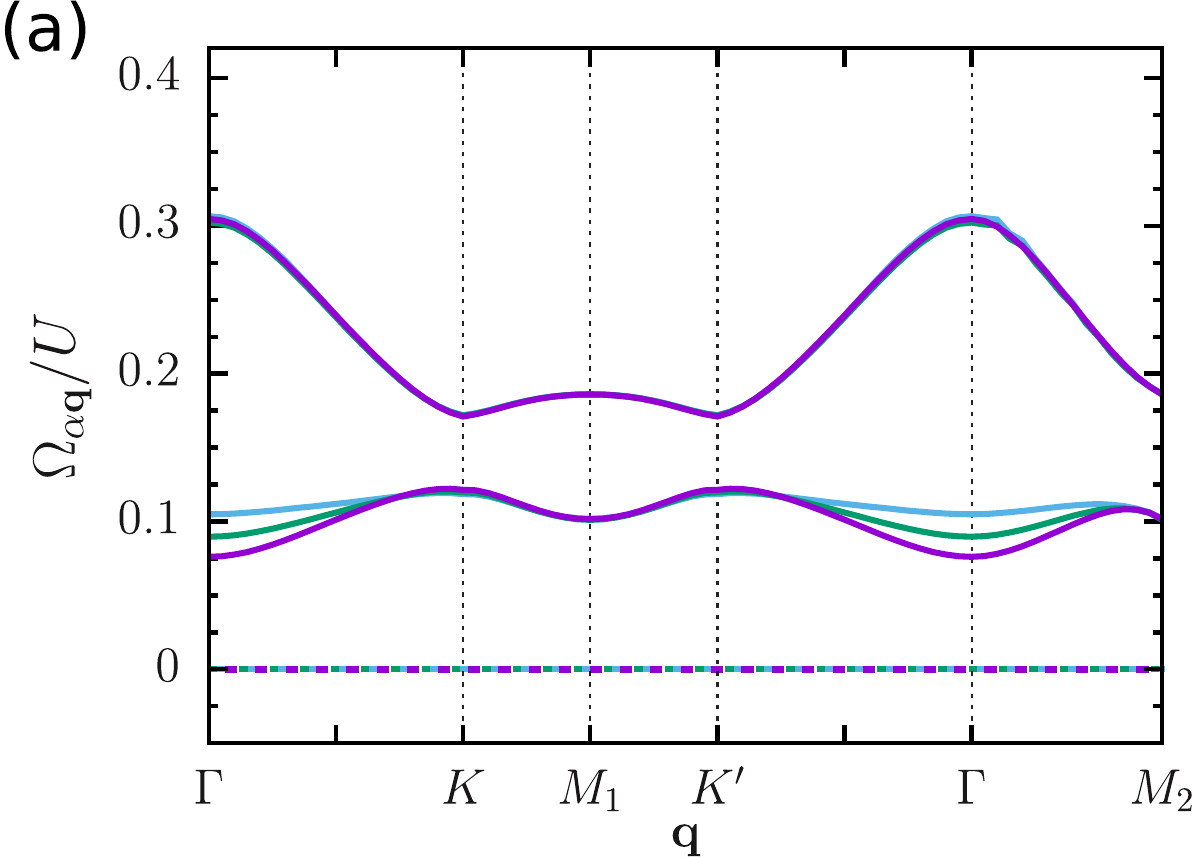}}
                   \vskip0.5cm 
\centerline{\includegraphics[width=5.5cm]{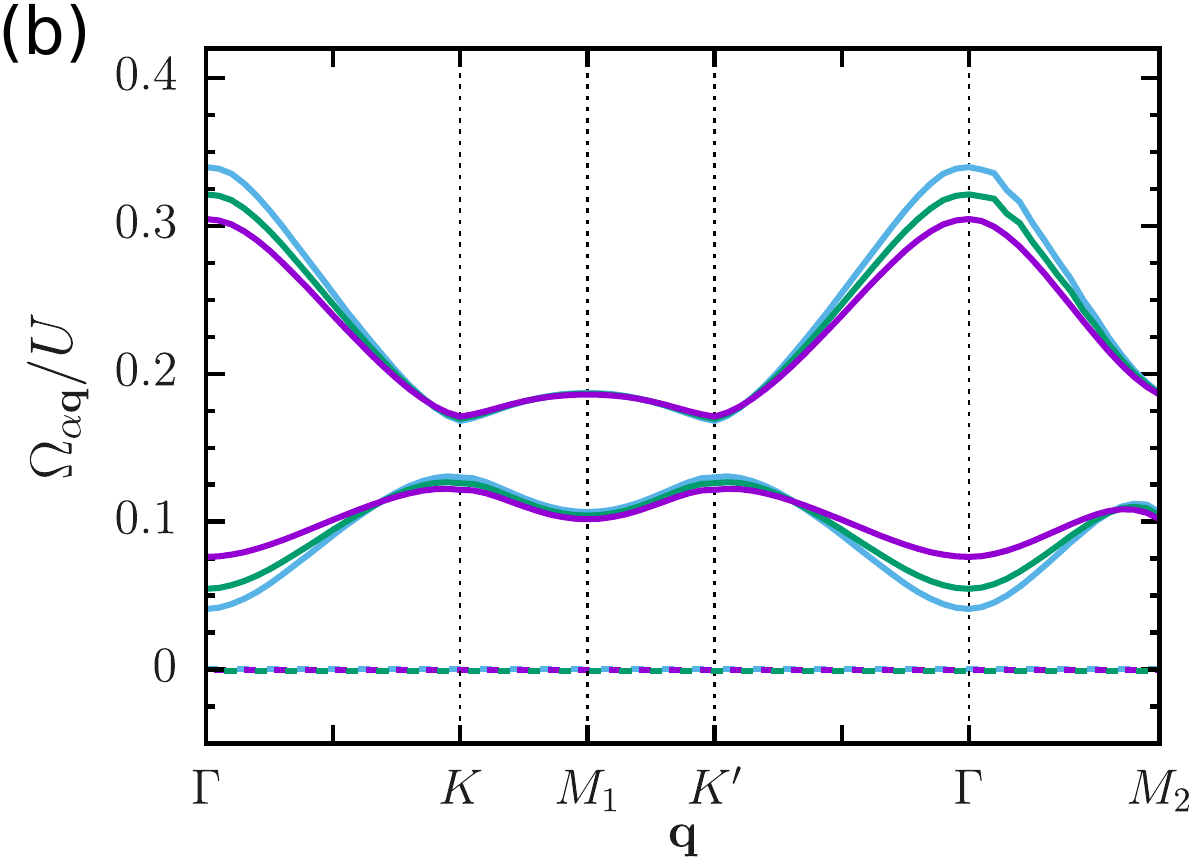}
}
\caption{SL excitations \eqref{eq:projS2}: Spin-wave spectrum \eqref{omega-b} 
along paths in the first BZ for on-site repulsion energies 
$U_A = U_B = U$ and the next-nearest-neighbor hopping amplitude
$t_2$ given by $\cos(\phi) = t_1/ (4 t_2)$.
Solid and dashed lines respectively represent the real part of $\Omega_{\pm,\bq}$
and the imaginary part of $\Omega_{+,\bq} = -\Omega_{-,\bq} $.
Phase
$\phi = 0.4$ [blue line in (a)], 
$\phi = 0.5$ [green line in (a)],  
$\phi = 0.656$ (magneta line),
$\phi = 0.7$ [green line in (b)], and  
$\phi = 0.8$ [blue line in (b)]. }   
\label{fig:espectro-phi}
\end{figure}

The dispersion relation \eqref{omega-b} 
[the spin-wave spectrum for the SL excitations \eqref{eq:projS2}]
for the nearly flat band limit \eqref{optimal-par} and on-site
repulsion energies $U_A = U_B = U$ is shown in Fig.~\ref{figEspectro}(d). 
Notice that, instead of the nearest-neighbor hopping energy $t_1$, 
the energy scale of the spin-wave spectrum is given by the
on-site repulsion energy $U$, since the kinetic coefficients 
\eqref{eq:omegaBar}  
(associated with the noninteracting bands $c$) are neglected.
Similarly to the ML excitations [Fig.~\ref{figEspectro}(a)], 
the spin-wave spectrum for the SL
excitations is also gapped and has two branches:
the gap of the lower branch is at the $\Gamma$ point of 
the first BZ while the gap of 
upper one is at the $K$ and $K'$ points.
In contrast with the correlated Chern insulator \cite{leite2021},
whose spin-wave spectrum has a Goldstone mode at the $\Gamma$ point
related with a continuous SU(2) symmetry that is spontaneously broken, 
the flat-band FM phase of the
correlated topological insulator \eqref{eqHH} has a gapped spectrum:
such a feature, that is properly described by the bosonization
formalism, is due to the fact that both the Hamiltonian \eqref{eqHH}
and the ground state \eqref{eq:FM} preserve a U(1) spin rotation symmetry
(see Sec.~II.A from Ref~\cite{doretto2015flat} 
and Ref.~\cite{neupert2012topological} for more details).
Differently from the corresponding correlated Chern insulator \cite{leite2021},
whose spin-wave spectrum has Dirac points at the $K$ and $K'$ points,
here one finds an energy gap between the lower and upper bands at the
$K$ and $K'$ points,
\begin{equation}
  \Delta^{(K)} = \Omega_{+,K} - \Omega_{-,K} = 4.96 \times 10^{-2}\, U;
\label{gap-SL}
\end{equation}
such a gap is large than the one [Eq.~\eqref{gap-ML}] found for the ML excitations.
Interestingly, apart from the energy gaps at the $\Gamma$, $K$, and
$K'$ points, the spin-wave spectrum shown in Fig.~\ref{figEspectro}(d)
qualitatively resembles the one of the correlated Chern
insulator on the honeycomb lattice that we have previously studied
(see Fig.~6(a) from Ref.~\cite{leite2021}). 
Finally, since the quadratic boson term \eqref{H42} is Hermitian,
the spin-wave excitations \eqref{omega-b} are real quantities, i.e.,
the decay rates of the spin-wave excitations 
vanish, in contrast with the behaviour of the ML excitations, which
display a quite small decay rate [see Figs.~\ref{figEspectro}(a) and (d)].
Importantly, for each momentum within the first BZ,  
the excitation energy associated with the upper band of the SL case
is lower than the corresponding value of the ML case, 
a feature also found for the lower bands 
[see Figs.~\ref{figEspectro}(a) and (d)]. Therefore, 
the SL excitations are indeed the lowest-energy excitations  
that characterize the flat-band FM
phase of the THM  \eqref{eqHH}, a feature that contrasts
with the square lattice correlated Z$_2$ topological insulator   
\cite{doretto2015flat}, whose elementary excitations of the
corresponding flat-band FM phase are of the ML type.

In addition to the THM \eqref{eqHH} with
homogeneous on-site repulsion energies $U_A = U_B = U$, the spin-wave spectrum 
with a sublattice dependent on-site energy $U_a$ was also determined.
We show the spin-wave spectrum \eqref{omega-b} 
for the nearly flat-band limit \eqref{optimal-par} and 
$U_B = 0.8\, U_A = 0.8\, U$ and 
$U_B = 0.6\, U_A = 0.6\, U$ in Figs.~\ref{figEspectro}(e) and (f), respectively.
Similarly to the ML excitations [Figs.~\ref{figEspectro}(b) and (c)], 
we find that a finite $\Delta U = U_A - U_B$ modifies the spin-wave spectrum 
as compare to the homogeneous case $U_A = U_B = U$.
In particular, it breaks the symmetry at the $K$ and $K'$ points 
displayed by the spin-wave spectrum in the homogeneous case.
Such an asymmetry at the $K$ and $K'$ points of the
spin-wave spectrum as $\Delta U$ increases was also found for 
the correlated Chern insulator \cite{leite2021}
and  it might be related to the fact that a Hubbard
term with $U_A \not= U_B$ breaks inversion symmetry.
Notice that, as the diference $\Delta U$ increases: 
The energies of the spin-wave excitations decrease; 
the energy gap between the lower and upper bands at the $K$ point decreases, 
\begin{align}
    \Delta^{(K)} &= 1.82 \times 10^{-2}\, U
    \quad {\rm for} \quad \Delta U = 0.2\,U,
\nonumber \\
    \Delta^{(K)} &= 1.34 \times 10^{-2}\, U
    \quad {\rm for} \quad \Delta U = 0.4\,U,
\nonumber
\end{align}
while the one at the $K'$ point increases,
\begin{align}
    \Delta^{(K')} &= 7.11 \times 10^{-2}\, U
    \quad {\rm for} \quad \Delta U = 0.2\,U,
\nonumber \\
    \Delta^{(K')} &= 9.26 \times 10^{-2}\, U
    \quad {\rm for} \quad \Delta U = 0.4\,U.
\nonumber
\end{align}
For $U_B > U_A$ (not shown here), similar features are observed, but now the energy
gap at the $K$ point increases instead of the one at $K'$ point.  
Again, similarly to the homogeneous case, the energies of spin-wave
spectrum of the SL case are lower than the corresponding ones of
the ML case for a fixed $\Delta U$.

We also investigate how the spin-wave spectrum \eqref{omega-b}
modifies as the THM \eqref{eqHH} is tuned away from
the nearly flat band limit \eqref{optimal-par},  once 
the next-nearest-neighbhor hopping amplitude $t_2$ and  
the phase $\phi$ are modified while the on-site Hubbard energies $U_a$
are kept fixed. As mentioned in Sec.~\ref{sec:Diagonalization}  
(see also Fig.~2 from Ref.~\cite{leite2021}),
the noninteracting electronic bands $c$ [Eq.~\eqref{eq:omega}] 
become more dispersive as the model \eqref{eqHH} moves away from the
nearly flat band limit \eqref{optimal-par}. 
In the following, we describe the effects on the spin-wave spectrum
only due to variations of the parameters $t_2$ and $\phi$.
We refer the reader to Appendix~\ref{sec:mass-term}  for a similar
discussion concerning the effects of a finite staggered on-site energy term 
in the Hamiltonian \eqref{eqHH}.

In Fig.~\ref{fig:espectro-phi}(a), it is shown
the spin-wave spectrum \eqref{omega-b} for $\phi = 0.656$, $0.7$, and $0.8$, 
hopping amplitude $t_2$ determined by $\cos(\phi) = t_1/(4 t_2)$, 
and on-site repulsion energies $U_A = U_B = U$.
We find that the spin-wave spectrum
(in units of the on-site Hubbard energy $U$) for $\phi = 0.7$ and $0.8$ is
rather similar to the one for the nearly-flat band limit \eqref{optimal-par},
which corresponds to $\phi = 0.656$.
As the parameter $\phi$ increases, one sees that only
the excitation energies of the lower band in the vicinity of the 
$\Gamma$ point increases while the rest of the spectrum remains almost
the same as compared with the one obtained for the nearly flat band limit
\eqref{optimal-par}. 
Indeed, for $\phi=0.8$, the energy gap of the lower band moves from
the $\Gamma$ to the $M_i$ points.
On the other hand, as shown in Fig.~\ref{fig:espectro-phi}(b), a
decreasing of the parameter $\phi$ from $\phi = 0.656$ yields:
A decreasing of the excitation energies of the lower band in the
vicinity of the  $\Gamma$ point;
an increasing of the excitation energies of the upper band around the same point; 
and a small decreasing in the energy gap between the lower and upper
bands at the $K$ and $K'$ points. 
Indeed, one finds that such energy gap
$\Delta^{(K)} = \Omega_{+,K} - \Omega_{-,K} = 3.81 \times 10^{-2}\, U$ ($\phi =0.4$),  
$4.35 \times 10^{-2}\, U$ ($\phi =0.5$),  
$4.96 \times 10^{-2}\, U$ ($\phi =0.656$),
$5.19 \times 10^{-2}\, U$ ($\phi =0.7$), and
$5.34 \times 10^{-2}\, U$ ($\phi =0.8$).
We believe that such rather small modifications in the spin-wave
spectrum as the model \eqref{eqHH} is tuned away from the nearly flat
band limit \eqref{optimal-par} might be due to the fact that
the main effects associated with the dispersion of the lower
noninteracting band $c$, that are encoded in the kinetic coefficients
\eqref{eq:omegaBar}, are not properly taken into account by the
bosonization scheme.

\begin{table}[t]
\centering
\caption{Chern numbers of the lower spin-wave bands \eqref{omega-b}
  for both the ML ($C_{ML}$) and the SL ($C_{SL}$) excitations
  at the nearly flat band limit \eqref{optimal-par}.} 
\begin{tabular}{lcccc}
\hline\hline
   &  & $U_A = U_B = U$  & &  $U_B = 0.8\, U_A = 0.8\, U$ \\ 
\hline 
   $C_{ML}$  &\quad \quad\quad\quad &  $\pm 0.29$  & \quad\quad\quad &  $\pm 0.18$  \\ 
   $C_{SL}$   &\quad \quad\quad\quad &  $\pm 1.17$  & \quad\quad\quad &  $\pm 1.08$ \\ 
\hline\hline
\end{tabular} 
\label{tb-chern}
\end{table}

Concerning the topological properties of the spin-wave bands, 
we find some evidences that the spin-wave bands for the SL excitations 
\eqref{eq:projS2} might be topologically nontrivial. In Table~\ref{tb-chern}, 
we present the Chern numbers $C_{SL}$ of the lower spin-wave bands \eqref{omega-b} 
for the SL excitations shown in Figs.~\ref{figEspectro}(d) and (e)
[the corresponding Chern numbers $C_{ML}$ for the ML excitations 
shown in Figs.~\ref{figEspectro}(a) and (b) are also included for comparison].
Such a feature contrasts with the one found for the corresponding
correlated Chern insulator on a honeycomb lattice
\cite{gu2019itinerant,leite2021}, whose spin-wave 
bands are topologically trivial in the completely flat band limit.
For more details about the topological properties of the spin-wave bands, 
we refer the reader to Appendix~\ref{sec:chernNumber2}.

%%%%%%%%%%%%%%%%%%%%%%%%%%%%%%%%%%%%%%%%%%%%%%%%%%%%%%%%%%%%%%%%%%%%%%%%%%%%%%%%%%%%%
\section{Summary}
\label{sec:summary}

In summary, in this paper we studied the flat-band FM
phase of a correlated Z$_2$ topological insulator on a honeycomb
lattice described by a topological Hubbard model, whose noninteracting
limit is given by a generalization of the spinless Haldane model
\cite{haldane1988model}.  
Such a study complements our previous one \cite{leite2021} concerning
the flat-band FM phase of a correlated Chern insulator
described by a Haldane-Hubbard model. 
We considered the model at $1/4$ filling of its noninteracting limit
and study the system  within a bosonization scheme for flat-band
correlated Z$_2$ topological insulators.
Our main result [Figs.~\ref{figEspectro}(d)] is the
calculation of the spin-wave excitation  
spectrum for the nearly flat band limit \eqref{optimal-par} of the
noninteracting lower bands and equal on-site repulsion
energies associated with the sublattices $A$ and $B$  
($U_A =U_B = U$).
Moreover, we also determined the spin-wave spectrum  
when an offset in the on-site repulsion energies is
introduced ($U_A \not= U_B$), and when the width of the lower
noninteracting bands increases due to changes in the parameters of the
noninteracting electronic Hamiltonian.

Differently from the correlated Chern insulator \cite{leite2021},
for the correlated topological insulator \eqref{eqHH}, one can, 
in principle, define two sets of boson
operators $b_0$ and $b_1$ as done in Eq.~\eqref{eq:bosons} considering
both the spin-flip excitations \eqref{eq:projS1}, that changes the
sublattice index (ML excitations), 
and the spin-flip excitations \eqref{eq:projS2}, that preserves the
sublattice index (SL excitations). 
We found that the spin-wave spectrum for both ML and SL
excitations are gapped and have two branches, with an energy gap 
between the lower and upper bands at the $K$ and $K'$ points of the
first BZ. Such features are in contrast with the ones found for 
the correlated Chern insulator on a honeycomb lattice
\cite{leite2021}, whose spin-wave spectrum has a 
Goldstone mode at the center of the BZ  
($\Gamma$ point) and Dirac points at the $K$ and $K'$ points.
Mostly important, the lowest-energy excitations are the SL ones, a
feature that is distinct from the one found for the square lattice
$\pi$-flux model \cite{doretto2015flat}, whose flat-band FM
phase is characterized by ML excitations: while both correlated Chern
insulators on the square \cite{doretto2015flat} and honeycomb
\cite{leite2021} lattices are characterize by the SL excitations, such
a common feature seems to be not shared by the corresponding
topological insulators.
Finally, our findings indicated that the spin-wave 
bands for the SL excitations might be topologically nontrivial, even 
in the completely flat band limit, a feature that also contrasts with
the behaviour of the corresponding correlated Chern insulator \cite{gu2019itinerant}.

%%%%%%%%%%%%%%%%%%%%%%%%%%%%%%%%%%%%%%%%%%%%%%%%%%%%%%%%%%%%%%%%%%%%%%%%%%%%%%%%%%%%%
\acknowledgments

We thank E. Miranda for helpful discussions and 
L.S.G.L. kindly acknowledges the financial support of the Conselho
Nacional de Desenvolvimento Cient\'ifico e Tecnol\'ogico (CNPq) under
the Grant No.~162323/2017-4.

\appendix
\begin{widetext}
%%%%%%%%%%%%%%%%%%%%%%%%%%%%%%%%%%%%%%%%%%%%%%%%%%%%%%%%%%%%%%%%%%%%%%%%%%%%%%%%%%%%%
\section{The $F_{\alpha \beta,\mathbf{q}}$ and 
            $\mathcal{G}_{\alpha \beta a \sigma}(\bk,\bq)$ functions
            for the ML excitations}
\label{sec:ap-details-mixed}

In this Appendix, the expansions of the $F_{\alpha \beta,\mathbf{q}}$
[Eq.~\eqref{eq:F2abhigh}] and 
the $\mathcal{G}_{\alpha \beta a \sigma}(\bk,\bq)$ functions
in terms of the coefficients \eqref{eqBs} are quoted. Such expressions were
previously derived by one of us in Ref.~\cite{doretto2015flat}.

From Eqs.~\eqref{eq:BogoTransf}, \eqref{eq:Bogocoef}, and \eqref{eq:g1}, 
one easily shows that Eq.~\eqref{eq:F2abhigh}  can be written as
\begin{align}
 F^2_{\alpha \beta, \bq }  =&  \frac{1}{4} \sum_\bp  
       \left[ (-1)^\alpha + (-1)^\beta\right] \left( 1 - \hat{B}_{3,\bp} \hat{B}_{3, -\bp+\bq} \right) 
     -\left[ (-1)^\alpha - (-1)^\beta\right]  \left( \hat{B}_{3,\bp} - \hat{B}_{3, -\bp + \bp} \right) 
\nonumber \\
    &+ \left[ 1 + (-1)^{\alpha+\beta}\right] \left( \hat{B}_{1,\bp} \hat{B}_{1, -\bp+\bq}   + \hat{B}_{2,\bp} \hat{B}_{2, -\bp+\bq} \right) 
      - i\left[ 1 - (-1)^{\alpha+\beta}\right] \left( \hat{B}_{1,\bp} \hat{B}_{2, -\bp+\bq}   - \hat{B}_{2,\bp} \hat{B}_{1, -\bp+\bq} \right), 
\label{eq:F2high}
\end{align}
where $\alpha,\beta = 0,1$ and $\hat{B}_{i,\bk} = B_{i,\bk} /|\mathbf{B}_\bk|$.
The $F^2_{\alpha \beta, \bq }$ function for the nearly flat band limit
\eqref{optimal-par} is shown in Fig.~\ref{fig:F2high}. It is clear
that the condition \eqref{conditionF} is not completely fulfilled by
the Haldane model \eqref{eqHH0}, since ${\rm Im}F^2_{01, \bq }$ and
${\rm Im}F^2_{10, \bq }$ are finite in the vicinity of the $M_1$ and
$M_2$ points. As discussed in Appendix~B from Ref.~\cite{leite2021}, 
in principle, such a result indicates that it is not possible to
define the two sets of independent boson operators $b_0$ and $b_1$ as
done in Eq.~\eqref{eq:bosons}, a feature that distinguishes the
Haldane model \eqref{eqHH0} from the square lattice $\pi$-flux model \cite{doretto2015flat}.
Due to the similarities between the topological insulator
\eqref{eqHH0} and the Chern insulator \cite{leite2021} 
and the fact that the bosonization scheme provides reasonable results
for the correlated Chern insulator described by the Haldane-Hubbard
model, we follow the lines of Ref.~\cite{leite2021} and {\sl assume}
that, for the topological insulator \eqref{eqHH0}, the bosons
operators $b_0$ and $b_1$ can be defined by Eq.~\eqref{eq:bosons}  and
that they constitute two sets of independent boson operators.

\begin{figure*}[t]
\centerline{
  \includegraphics[width=5.5cm]{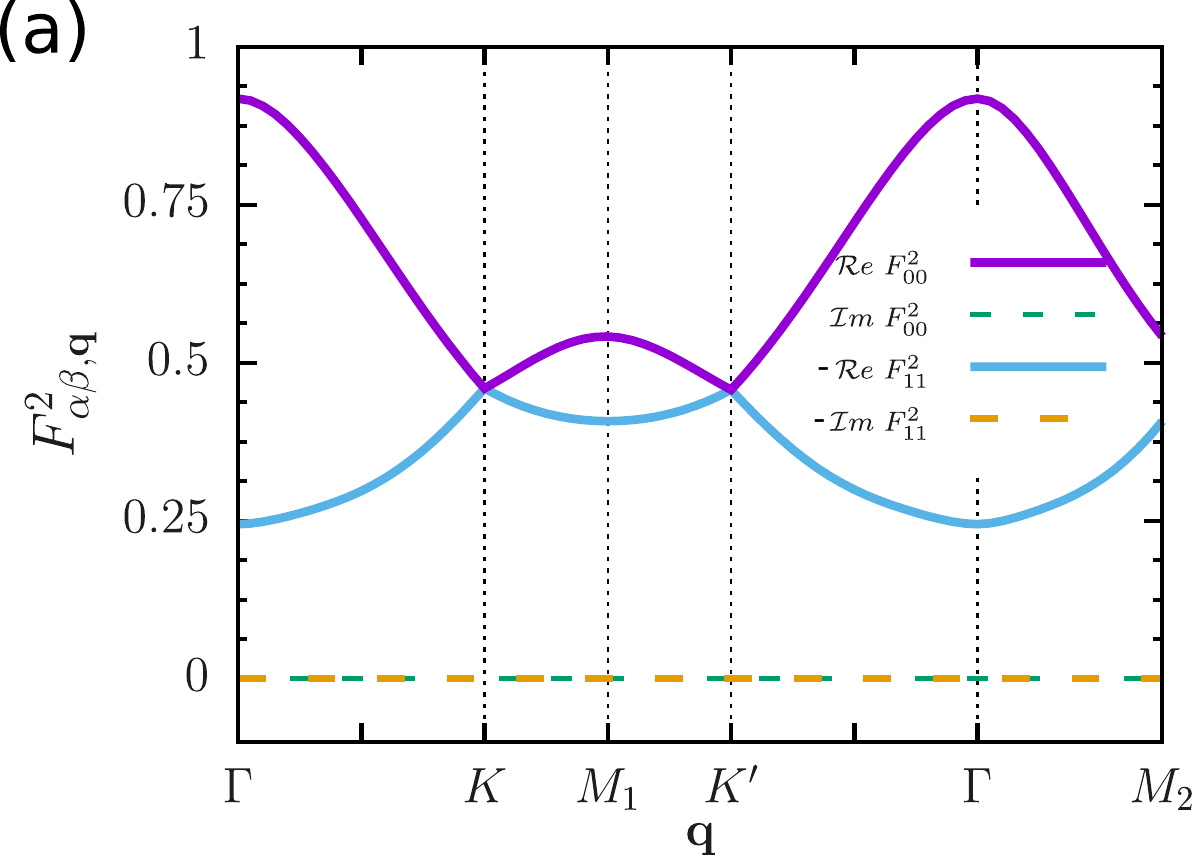} 
  \hskip1.0cm
  \includegraphics[width=5.5cm]{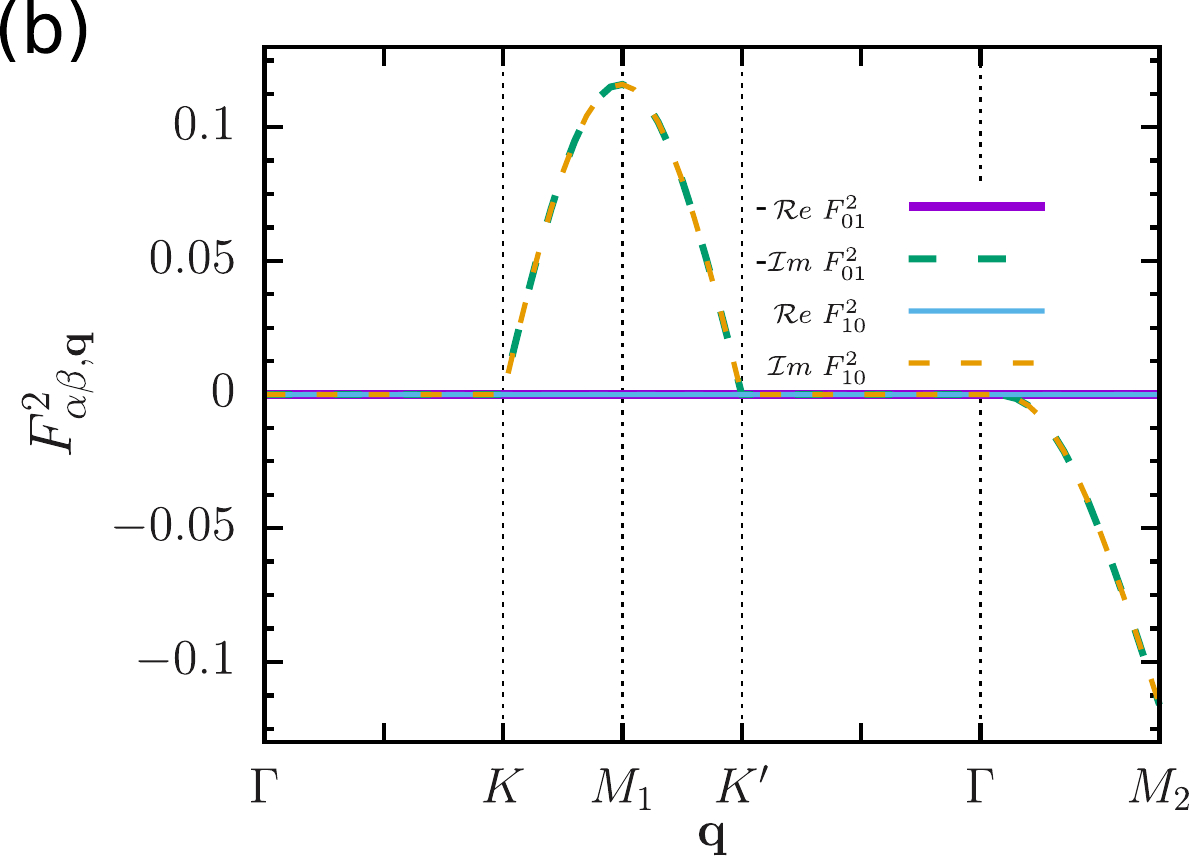}
}
\caption{ML excitations: The real (solid line) and imaginary (dashed line) parts 
 of $F_{\alpha \beta,\bq}^2$  
 [Eq.~\eqref{eq:F2abhigh}] for the
 Haldane model \eqref{eqHH0} in the 
 nearly-flat band limit \eqref{optimal-par}
 along paths in the first BZ:  
 (a) $F_{00,\bq}^2$ and $-F_{11,\bq}^2$ and 
 (b) $F_{01,\bq}^2$ and $F_{10,\bq}^2$.}
\label{fig:F2high}
\end{figure*}

Once the expansion of the $F^2_{\alpha \beta, \bq }$ function in terms
of the coefficients \eqref{eqBs} is known, 
one can easily determine the kinetic
coefficients \eqref{eq:omegaBar} [compare the integrands of
Eqs.~\eqref{eq:F2abhigh} and \eqref{eq:omegaBar}]. 
For instance,  in Figs.~\ref{fig:omegabar-high}(a) and (b), one shows the kinetic
coefficients \eqref{eq:omegaBar} for the nearly flat band limit
\eqref{optimal-par}.

\begin{figure*}[b]
\centerline{
 \includegraphics[width=5.5cm]{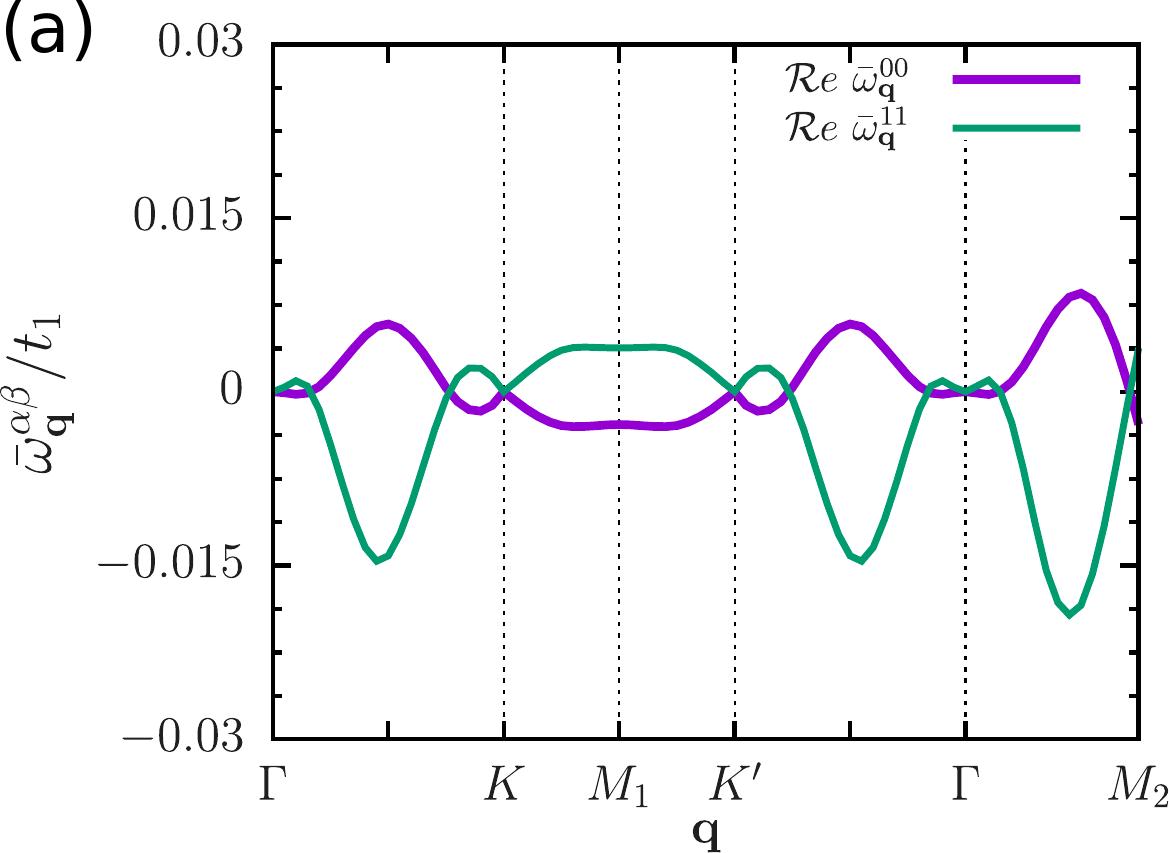}
 \hskip0.5cm
 \includegraphics[width=5.5cm]{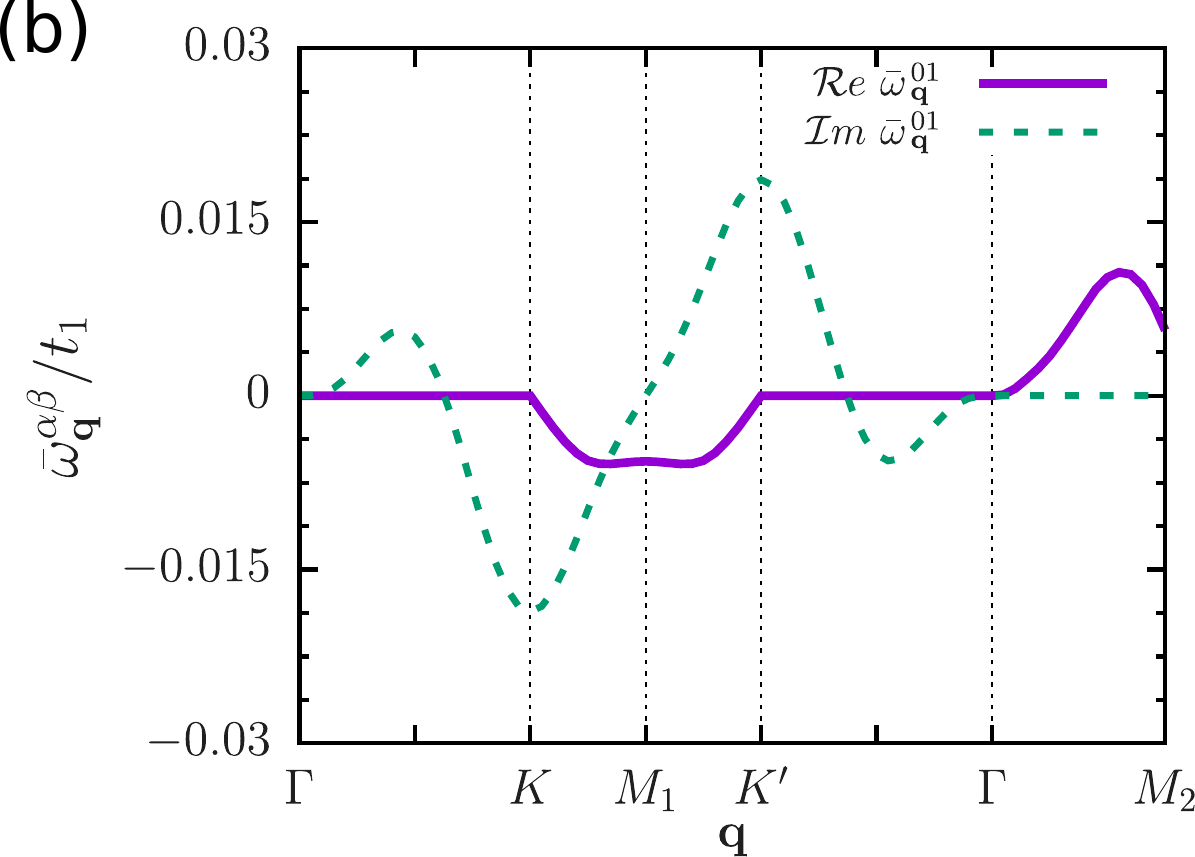}
 \hskip0.5cm
 \includegraphics[width=5.5cm]{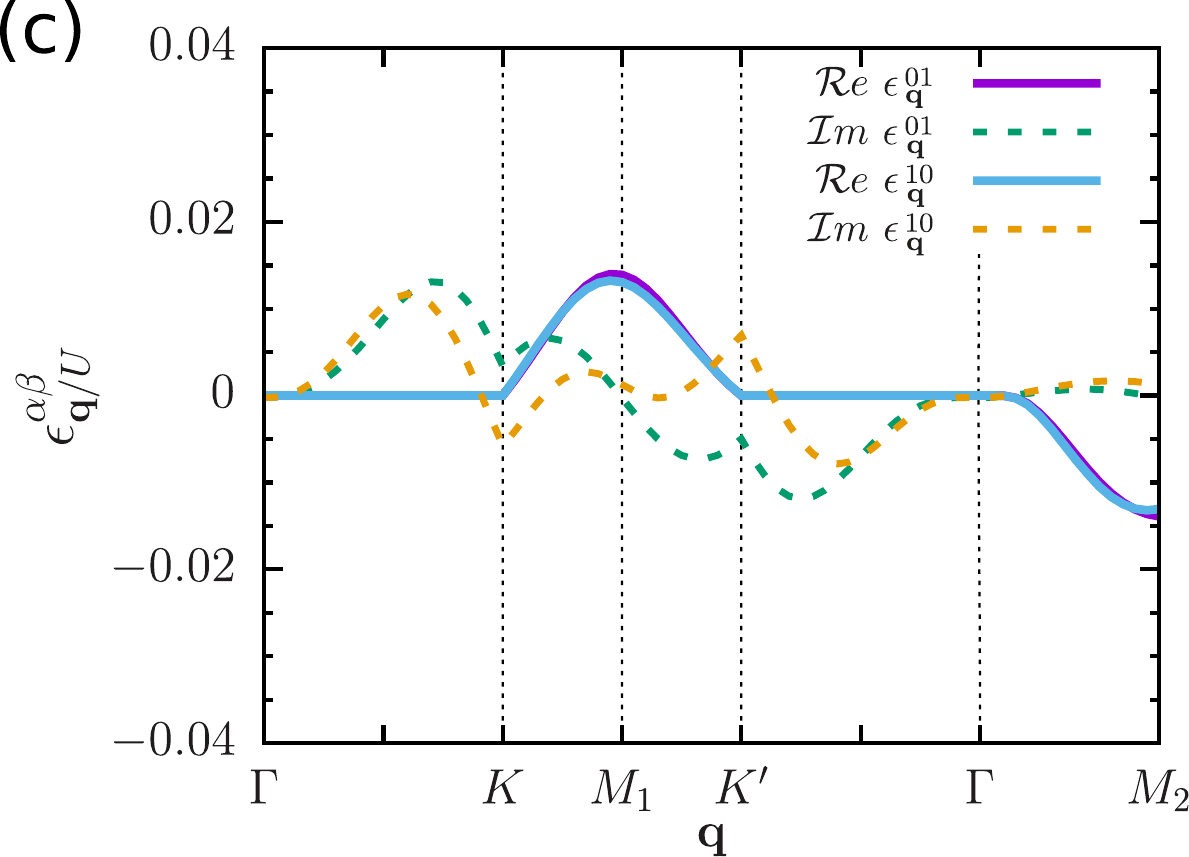}
}
\caption{ML excitations: The real (solid line) and imaginary (dashed line) parts of 
 the kinetic coefficients 
 (a) $\bar{\omega}^{00}_\bq$  and $\bar{\omega}^{11}_\bq$ and
 (b) $\bar{\omega}^{01}_\bq$  
 [Eq.~\eqref{eq:omegaBar}] along paths in the first BZ  
 for the Haldane model \eqref{eqHH0} in the nearly-flat band
 limit \eqref{optimal-par}.
 (c) The real (solid line) and imaginary (dashed line) parts of the 
 coefficients $\epsilon^{01}_\bq$ and $\epsilon^{10}_\bq$  
 [Eq.~\eqref{eq:Epsilon}]  for the THM  
 \eqref{eqHH} in the nearly-flat band
 limit \eqref{optimal-par} and on-site repulsion energies $U_A = U_B = U$.}
\label{fig:omegabar-high}
\end{figure*}

The $\mathcal{G}_{\alpha \beta a \sigma}(\bk,\bq)$ function is defined as
\begin{eqnarray}
  \mathcal{G}_{\alpha \beta a \uparrow}( \mathbf{k} , \mathbf{q}) &=& 
   -\sum_{ \mathbf{p}} 
    \frac {G_{a\,\uparrow}(\mathbf{p}, \mathbf{k})}{F_{\alpha\alpha, \mathbf{q}} F_{\beta\beta, \mathbf{k+q}}}
    g_{\alpha}(\mathbf{p-k}, \mathbf{q})
    g_{\beta}^*(-\bp+\bk+\bq,\bk+\bq),
\nonumber \\
   \mathcal{G}_{\alpha \beta a \downarrow}( \mathbf{k} , \mathbf{q}) &=& 
   + \sum_{ \mathbf{p}} 
    \frac {G_{a\,\downarrow}(\bp - \bq, \bk)} {F_{\alpha\alpha, \mathbf{q}} F_{\beta\beta, \mathbf{k+q}}}
    g_{\alpha}(\mathbf{p}, \mathbf{q})
    g_{\beta}^*(-\bp+\bk+\bq,\bk+\bq),
\label{Gcal} 
\end{eqnarray} 
where $G_{a\,\sigma}(\bp,\bq)$ is given by Eq.~\eqref{eq:ga-sigma}.  
With the aid of Eq.~\eqref{eq:Bogocoef}, one finds that
\begin{align}
\mathcal{G}_{\alpha \beta a \sigma }( \bk , \bq) = 
  &-\gamma_\sigma\frac{1}{8} \left[\delta_{a,A} + \delta_{a,B}(-1)^{\alpha + \beta} \right] 
                   \frac{1}{F_{\alpha\alpha, \mathbf{q}} F_{\beta\beta, \mathbf{k+q}}} 
   \nonumber \\
  & \times  \sum_\bp  
     \zeta_1(\sigma) \left[1 + \gamma_\sigma(-1)^a \hat{B}_3(1)\right]  
                               \left[1 - \gamma_\sigma(-1)^a \hat{B}_3(2)\right]  \left[1 - (-1)^a \hat{B}_3(3)\right]   
\nonumber \\
  &+\zeta_2(\sigma)   \left[ \hat{B}_1(2) \hat{B}_1(3) + \hat{B}_2(2) \hat{B}_2(3)   
           + i(-1)^a \left( \hat{B}_1(2) \hat{B}_2(3) -\hat{B}_2(2) \hat{B}_1(3) \right)\right] 
           \left[ 1 -(-1)^{a} \hat{B}_3(1)\right]
\nonumber \\
  &+ \zeta_3(\sigma)  \left[ \hat{B}_1(1) \hat{B}_1(3) + \hat{B}_2(1) \hat{B}_2(3)   
           + i(-1)^a \left( \hat{B}_1(1) \hat{B}_2(3) -\hat{B}_2(1) \hat{B}_1(3) \right)\right] 
           \left[ 1 -(-1)^{a} \hat{B}_3(2)\right]
\nonumber \\
  &+ \zeta_4(\sigma)  \left[ \hat{B}_1(1) \hat{B}_1(2) + \hat{B}_2(1) \hat{B}_2(2)   
      + i\gamma_\sigma(-1)^a \left( \hat{B}_2(1) \hat{B}_1(2) -\hat{B}_1(1) \hat{B}_2(2) \right)\right] 
           \left[ 1 -(-1)^{a} \hat{B}_3(3)\right],
\label{Gcal2}
\end{align}
where the coefficient $\gamma_\uparrow = -\gamma_\downarrow = 1$, the
coefficients $\zeta_i(\sigma)$ read
\begin{align}
 \zeta_1(\uparrow) &= (-1)^\alpha, \quad\quad
 \zeta_2(\uparrow) = (-1)^\beta, \quad\quad
 \zeta_3(\uparrow) = 1, \quad\quad
 \zeta_4(\uparrow) = (-1)^{\alpha+\beta}, 
\nonumber \\
 \zeta_1(\downarrow) &= (-1)^\beta, \quad\quad
 \zeta_2(\downarrow) = (-1)^{\alpha+\beta},\quad\quad 
 \zeta_3(\downarrow) = (-1)^\alpha, \quad\quad
 \zeta_4(\downarrow) = 1, 
\end{align}
and the $\hat{B}_i(j)$ functions, with $i,j = 1,2,3$, are given by
\begin{align}
 \hat{B}_i(1) &= \hat{B}_{i,-\bp+\bk+\bq}, \quad\quad
 \hat{B}_i(2) = \hat{B}_{i,\bp}, \quad\quad
 \hat{B}_i(3) = \hat{B}_{i,+\bp-\bk}, 
  \quad\quad {\rm for} \quad \sigma = \,\uparrow, 
\nonumber \\
 \hat{B}_i(1) &= \hat{B}_{i,-\bp+\bk+\bq}, \quad\quad
 \hat{B}_i(2) = \hat{B}_{i,\bp}, \quad\quad
 \hat{B}_i(3) = \hat{B}_{i,-\bp+\bq}, 
\label{Bidown-ml}
  \quad\quad {\rm for} \quad \sigma = \,\downarrow.
\end{align}

Equations~\eqref{eq:F2high} and \eqref{Gcal2} allow us to determine
the coefficients \eqref{eq:Epsilon}.
In particular, the coefficients 
$\epsilon^{01}_\bq$ and $\epsilon^{10}_\bq$ 
for the nearly flat band limit \eqref{optimal-par} and on-site
repulsion energies $U_A = U_B = U$ are plotted in 
Fig.~\ref{fig:omegabar-high}(c). One sees that
$\epsilon^{01}_\bq \not= (\epsilon^{10}_\bq)^*$, which implies that
the quadratic bosonic Hamiltonian \eqref{H42} is non-Hermitian for the
ML excitations \eqref{eq:projS1}.

%%%%%%%%%%%%%%%%%%%%%%%%%%%%%%%%%%%%%%%%%%%%%%%%%%%%%%%%%%%%%%%%%%%%%%%%%%%%%%%%%%%%%
\section{The $F_{\alpha \beta,\mathbf{q}}$ and 
            $\mathcal{G}_{\alpha \beta a \sigma}(\bk,\bq)$ functions
            for the SL excitations}
\label{sec:ap-details-same}

In this Appendix, we present the equivalent of  Eqs.~\eqref{eq:F2high}--\eqref{Bidown-ml}
for the SL excitations \eqref{eq:projS2}.  
Indeed, such kind of spin-flip excitations were considered in
Ref.~\cite{leite2021} in the description of the flat-band
FM phase of a correlated Chern insulator described by a
Haldane-Hubbard model. However, since the canonical transformation 
\eqref{eq:BogoTransf} differs from the one employed in the study of
the correlated Chern insulator (see Eq.~(13) from Ref.~\cite{leite2021}),
the expressions of the $F_{\alpha \beta,\mathbf{q}}$ and
$\mathcal{G}_{\alpha \beta a \sigma}(\bk,\bq)$ functions are distinct
from the ones shown in Appendix A from Ref.~\cite{leite2021}.

From Eqs.~\eqref{eq:BogoTransf}, \eqref{eq:Bogocoef}, and \eqref{eq:g2}, 
one shows that, for the SL excitations, Eq.~\eqref{eq:F2abhigh} assumes the form
\begin{align}
F^2_{\alpha \beta, \bq }  =&  \frac{1}{4} \sum_\bp  
       \left[ 1 + (-1)^{\alpha+\beta}\right] \left( 1 + \hat{B}_{3,\bp} \hat{B}_{3, -\bp+\bq} \right) 
     + \left[ (-1)^\alpha + (-1)^\beta\right] \left( \hat{B}_{1,\bp} \hat{B}_{1, -\bp+\bq}   - \hat{B}_{2,\bp} \hat{B}_{2, -\bp+\bq} \right) 
\nonumber \\
    &-\left[ 1 - (-1)^{\alpha+\beta}\right]  \left( \hat{B}_{3,\bp} + \hat{B}_{3, -\bp + \bp} \right) 
      - i\left[ (-1)^\alpha - (-1)^\beta\right] \left( \hat{B}_{1,\bp} \hat{B}_{2, -\bp+\bq}   + \hat{B}_{2,\bp} \hat{B}_{1, -\bp+\bq} \right), 
\label{eq:F2low}
\end{align}
where $\alpha,\beta = 0,1$ and $\hat{B}_{i,\bk} = B_{i,\bk} /|\mathbf{B}_\bk|$. 
For the nearly flat band limit \eqref{optimal-par},
Eq.~\eqref{eq:F2low} is plotted in Figs.~\ref{fig:F2low}(a) and (b). 
Similarly to the ML excitations, one sees that ${\rm Im}F^2_{01, \bq }$ and
${\rm Im}F^2_{10, \bq }$ are finite in the vicinity of the $M_1$ and
$M_2$ points, implying that the condition \eqref{conditionF} is not
satisfied by all momenta in the first BZ.

\begin{figure*}[t]
\centerline{
  \includegraphics[width=5.5cm]{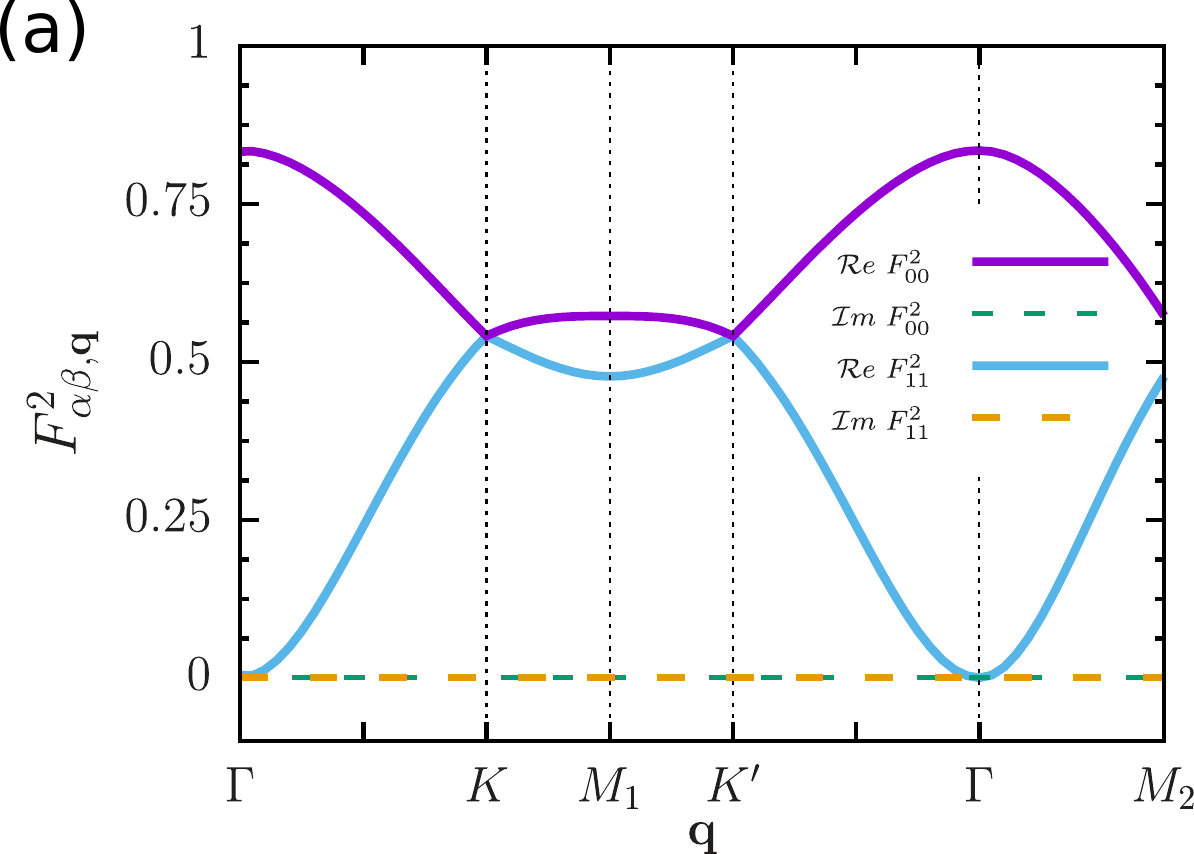} 
  \hskip0.5cm
  \includegraphics[width=5.5cm]{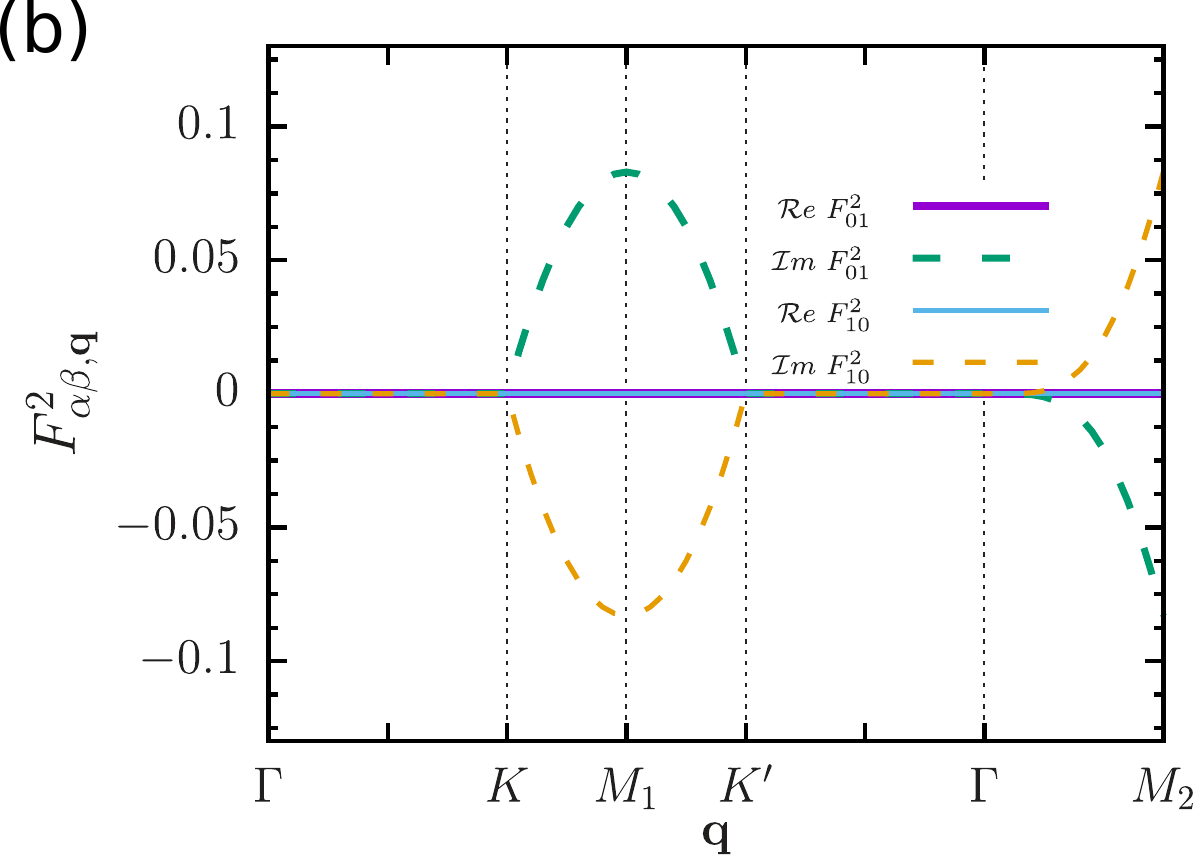}
  \hskip0.5cm
  \includegraphics[width=5.5cm]{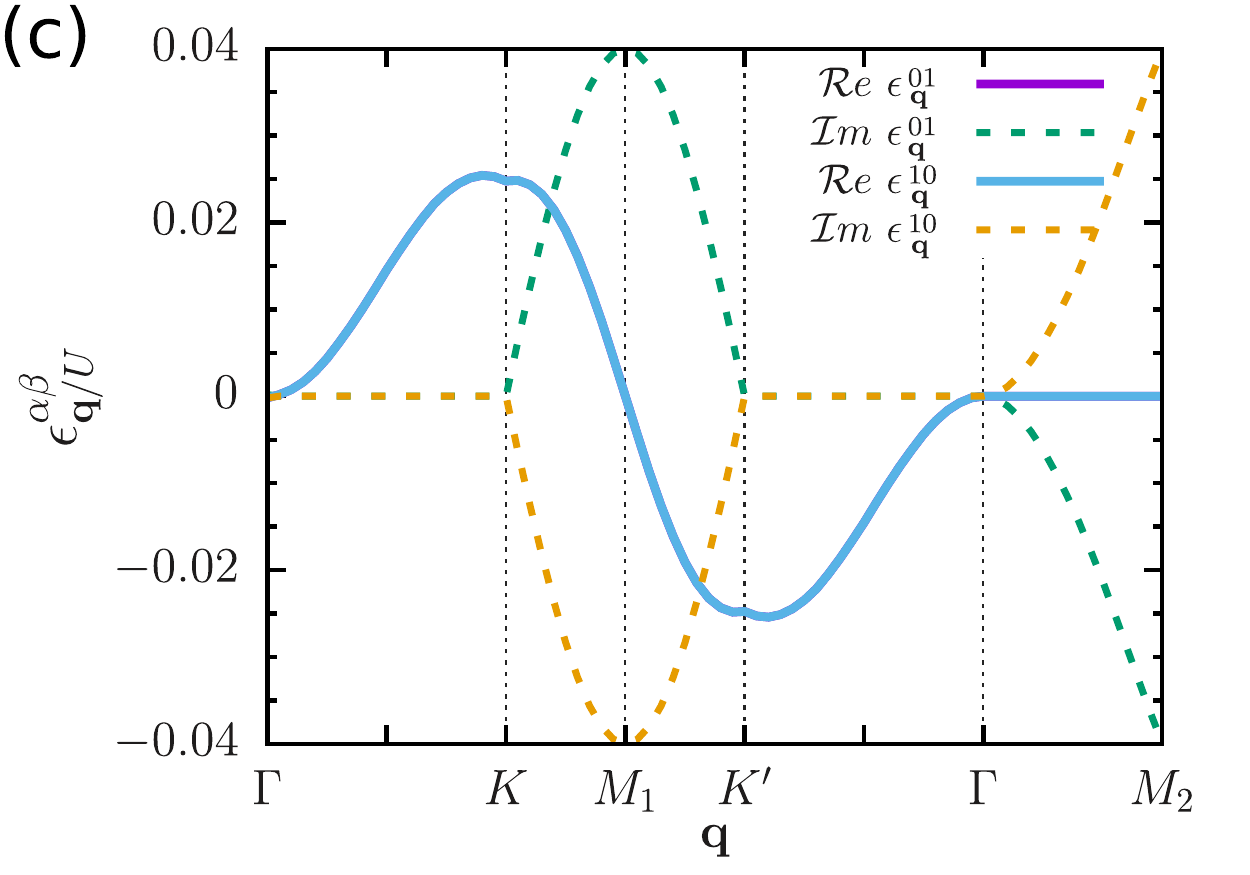}
} 
\caption{SL excitations: The real (solid line) and imaginary (dashed line) parts 
 of $F_{\alpha \beta,\bq}^2$  
 [Eq.~\eqref{eq:F2abhigh}] for the
 Haldane model \eqref{eqHH0} in the 
 nearly-flat band limit \eqref{optimal-par}
 along paths in the first BZ:  
 (a) $F_{00,\bq}^2$ and $F_{11,\bq}^2$ and 
 (b) $F_{01,\bq}^2$ and $F_{10,\bq}^2$. 
 (c) The real (solid line) and imaginary (dashed line) parts of the 
 coefficients $\epsilon^{01}_\bq$ and $\epsilon^{10}_\bq$  
 [Eq.~\eqref{eq:Epsilon}] 
 for the THM  \eqref{eqHH} in the nearly-flat band
 limit \eqref{optimal-par} and on-site repulsion energies $U_A = U_B = U$.}
\label{fig:F2low}
\end{figure*}

For the SL excitations, $\mathcal{G}_{\alpha \beta a \sigma}(\bk,\bq)$ 
is also defined by Eq.~\eqref{Gcal2}, but now it reads
\begin{align}
\mathcal{G}_{\alpha \beta a \sigma }( \bk , \bq) = 
  &-\gamma_\sigma\frac{1}{8} \left[\delta_{a,A} + \delta_{a,B}(-1)^{\alpha + \beta} \right] 
                   \frac{1}{F_{\alpha\alpha, \mathbf{q}} F_{\beta\beta, \mathbf{k+q}}} 
\nonumber \\
 & \times  
                  \sum_\bp  
                               \left[1 - (-1)^a \hat{B}_3(1)\right]  
                               \left[1 - (-1)^a \hat{B}_3(2)\right]  \left[1 - (-1)^a \hat{B}_3(3)\right]   
\nonumber \\
  &+\zeta_1(\sigma)   \left[ \hat{B}_1(2) \hat{B}_1(3) + \gamma_\sigma\hat{B}_2(2) \hat{B}_2(3)   
           + i(-1)^a \left( \hat{B}_1(2) \hat{B}_2(3) -\gamma_\sigma\hat{B}_2(2) \hat{B}_1(3) \right)\right] 
           \left[ 1 + \gamma_\sigma(-1)^{a} \hat{B}_3(1)\right]
\nonumber \\
  &+ \zeta_2(\sigma)  \left[ \hat{B}_1(1) \hat{B}_1(3) - \gamma_\sigma\hat{B}_2(1) \hat{B}_2(3)   
           + i(-1)^a \left( \hat{B}_1(1) \hat{B}_2(3) + \gamma_\sigma\hat{B}_2(1) \hat{B}_1(3) \right)\right] 
           \left[ 1 - \gamma_\sigma(-1)^{a} \hat{B}_3(2)\right]
\nonumber \\
  &+ \zeta_3(\sigma)  \left[ \hat{B}_1(1) \hat{B}_1(2) - \hat{B}_2(1) \hat{B}_2(2)   
      - i(-1)^a \left( \hat{B}_1(1) \hat{B}_2(2) +\hat{B}_2(1) \hat{B}_1(2) \right)\right] 
           \left[ 1 - (-1)^{a} \hat{B}_3(3)\right],
\label{Gcal3}
\end{align}
where the coefficient $\gamma_\uparrow = -\gamma_\downarrow = 1$, the
coefficients $\zeta_i(\sigma)$ are given by
\begin{align}
 \zeta_1(\uparrow) &= (-1)^{\alpha+\beta}, \quad\quad
 \zeta_2(\uparrow) = (-1)^\alpha, \quad\quad
 \zeta_3(\uparrow) = (-1)^\beta, 
\nonumber \\
 \zeta_1(\downarrow) &= (-1)^\alpha, \quad\quad
 \zeta_2(\downarrow) = (-1)^{\alpha+\beta}, \quad\quad
 \zeta_3(\downarrow) = (-1)^\beta, 
\end{align}
and the $\hat{B}_i(j)$ functions, with $i,j = 1,2,3$, are defined as
\begin{align}
 \hat{B}_i(1) &= \hat{B}_{i,-\bp+\bk+\bq}, \quad\quad
 \hat{B}_i(2) = \hat{B}_{i,\bp}, \quad\quad
 \hat{B}_i(3) = \hat{B}_{i,+\bp-\bk}, 
  \quad\quad {\rm for} \quad \sigma = \,\uparrow, 
\nonumber \\
 \hat{B}_i(1) &= \hat{B}_{i,-\bp+\bk+\bq}, \quad\quad
 \hat{B}_i(2) = \hat{B}_{i,\bp}, \quad\quad 
 \hat{B}_i(3) = \hat{B}_{i,-\bp+\bq}, 
\quad\quad {\rm for} \quad \sigma = \,\downarrow.
\end{align}

With the aid of Eqs.~\eqref{eq:F2low} and \eqref{Gcal3}, one can calculate
the coefficients \eqref{eq:Epsilon}.
For instance, the coefficients 
$\epsilon^{01}_\bq$ and $\epsilon^{10}_\bq$ 
for the nearly flat band limit \eqref{optimal-par} and on-site
repulsion energies $U_A = U_B = U$ are shown in 
Fig.~\ref{fig:F2low}(c).
Since $\epsilon^{01}_\bq = (\epsilon^{10}_\bq)^*$, 
the quadratic bosonic Hamiltonian \eqref{H42} is Hermitian for the
SL excitations \eqref{eq:projS2}.

\end{widetext}

%%%%%%%%%%%%%%%%%%%%%%%%%%%%%%%%%%%%%%%%%%%%%%%%%%%%%%%%%%%%%%%%%%%%%%%%%%%%%%%%%%%%% 
\section{Spin-wave spectrum for the ML excitations}
\label{sec:ap-spin-wave}

Here we discuss in details the behaviour of the spin-wave spectrum 
\eqref{omega-b}  for the ML excitations \eqref{eq:projS1}. 
In this case, one should consider the expressions of   
$g_\alpha(\bp, \bq)$,
$F_{\alpha\beta, \bq}$, and 
$\mathcal{G}_{\alpha \beta a \sigma}(\bp,\bq)$ 
respectively given by Eqs.~\eqref{eq:g1},
\eqref{eq:F2high}, and \eqref{Gcal2} in order to determine  
the kinetic coefficients \eqref{eq:omegaBar} 
and the coefficients \eqref{eq:Epsilon}.

Before discussing the behaviour of the spin-wave spectrum for the
ML excitations, a few remarks here about the dispersion relation
\eqref{omega-b} are in order:
(i) We follow the procedure adopted in our previous study \cite{leite2021} for the flat-band
FM phase of a correlated Chern insulator described by a
Haldane-Hubbard model, and completely neglect the contribution of the
kinetic coefficients \eqref{eq:omegaBar}; indeed, for the ML excitations, 
we find that $\bar{\omega}^{\alpha \alpha}_{\mathbf{q}}$ are
real while $\bar{\omega}^{01}_{\mathbf{q}}$ and $\bar{\omega}^{10}_{\mathbf{q}}$ 
are finite complex quantities, but rather small in units of the nearest-neighbor hopping
energy $t_1$ [see Figs.~\ref{fig:omegabar-high}(a) and (b)];
as discussed in detail in Ref.~\cite{leite2021}, we believe
that such finite values for $\bar{\omega}^{\alpha \beta}_{\mathbf{q}}$
are related to the symmetries of the Haldane model \eqref{eqHH0} and
to the fact that the condition \eqref{conditionF}
is not fulfilled for all momenta $\bq$ within the first BZ  
(see Fig.~\ref{fig:F2high}).
(ii) Concerning the coefficients \eqref{eq:Epsilon},  
we find that they are also complex quantities, with
$\epsilon^{\alpha \alpha}_\bq $ having a quite small imaginary part
and $\epsilon^{01}_\bq \not= (\epsilon^{10}_\bq)^*$ as shown in 
Fig.~\ref{fig:omegabar-high}(c);
such features imply that the quadratic Hamiltonian \eqref{H42}
is non-Hermitian, a behaviour previously found for the correlated
Chern insulator \cite{leite2021};
at the moment, we believe that the non-Hermiticity of the Hamiltonian
\eqref{H42} might be an artifact of the bosonization formalism
associated with the fact that the condition \eqref{conditionF} is not
completely satisfied by the Haldane model \eqref{eqHH0};
however, for the correlated Chern insulator \cite{leite2021}, 
the presence of the off-diagonal terms 
$(\alpha,\beta) = (0,1)$ and $(1,0)$ of the quadratic bosonic Hamiltonian \eqref{H42}
are indeed important, since they yield a spin-wave spectrum with 
Dirac points at the $K$ and $K'$ points of the first BZ  
(see Fig.~6 from Ref.~\cite{leite2021}),
in agreement with the numerical calculations \cite{gu2019itinerant};  
therefore, for the ML excitations, we also consider the complete and
non-Hermitian quadratic Hamiltonian \eqref{H42}.
For more details about these two important issues, 
we refer the reader to Sec.~VI.B and Appendix~B from Ref.~\cite{leite2021}.

Figure~\ref{figEspectro}(a) shows the dispersion relation \eqref{omega-b} 
for the nearly flat band limit \eqref{optimal-par} and on-site
repulsion energies $U_A = U_B = U$.
One sees that the spin-wave spectrum for the ML excitations is gapped
and has two branches: the gap of the lower branch is at the $M_i$ points of
the first BZ while the gap of the upper one is at the $K$ and $K'$ points.
Small energy gaps between the lower and upper bands at the $K$ and
$K'$ points are found, 
\begin{equation}
  \Delta^{(K)} = \Omega_{+,K} - \Omega_{-,K} = 1.04 \times 10^{-2}\, U,
\label{gap-ML}
\end{equation}
in contrast with the corresponding correlated Chern insulator \cite{leite2021},
whose spin-wave spectrum displays Dirac points at the $K$ and $K'$ points. 
Due to the non-Hermiticity of the quadratic boson term \eqref{H42},
one finds that the  the spin-wave excitations \eqref{omega-b} 
have a quite small decay rate (the imaginary part of
$\Omega_{\pm,\bq}$) along the $K$-$M_1$-$K'$ line, i.e., 
at the border of the first BZ 
[see the dashed line in Fig.~\ref{figEspectro}(a) and 
note the multiplicative factor 20]. Such a feature was also
found in the study of the correlated Chern insulator in
Ref.~\cite{leite2021}.

\begin{figure*}[t]
\centerline{
  \includegraphics[width=5.5cm]{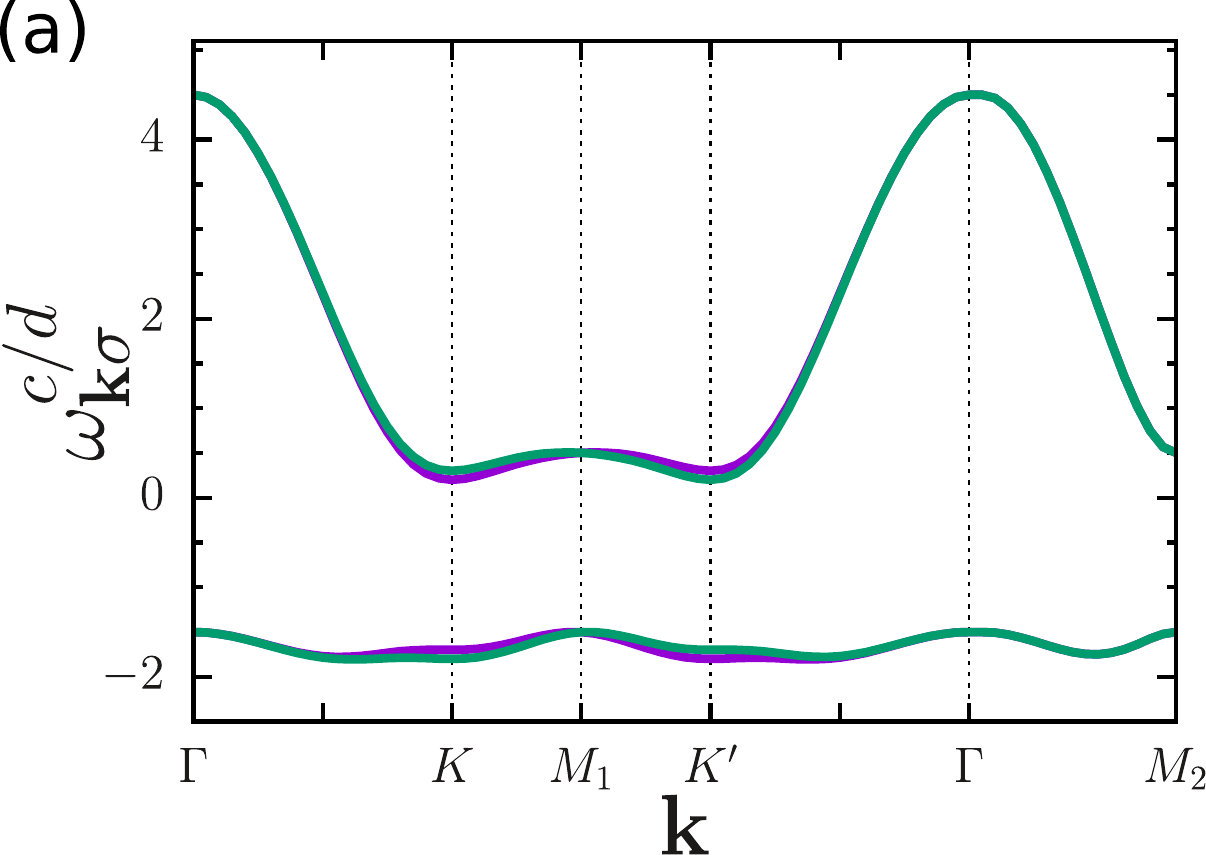}
\hskip0.5cm
  \includegraphics[width=5.5cm]{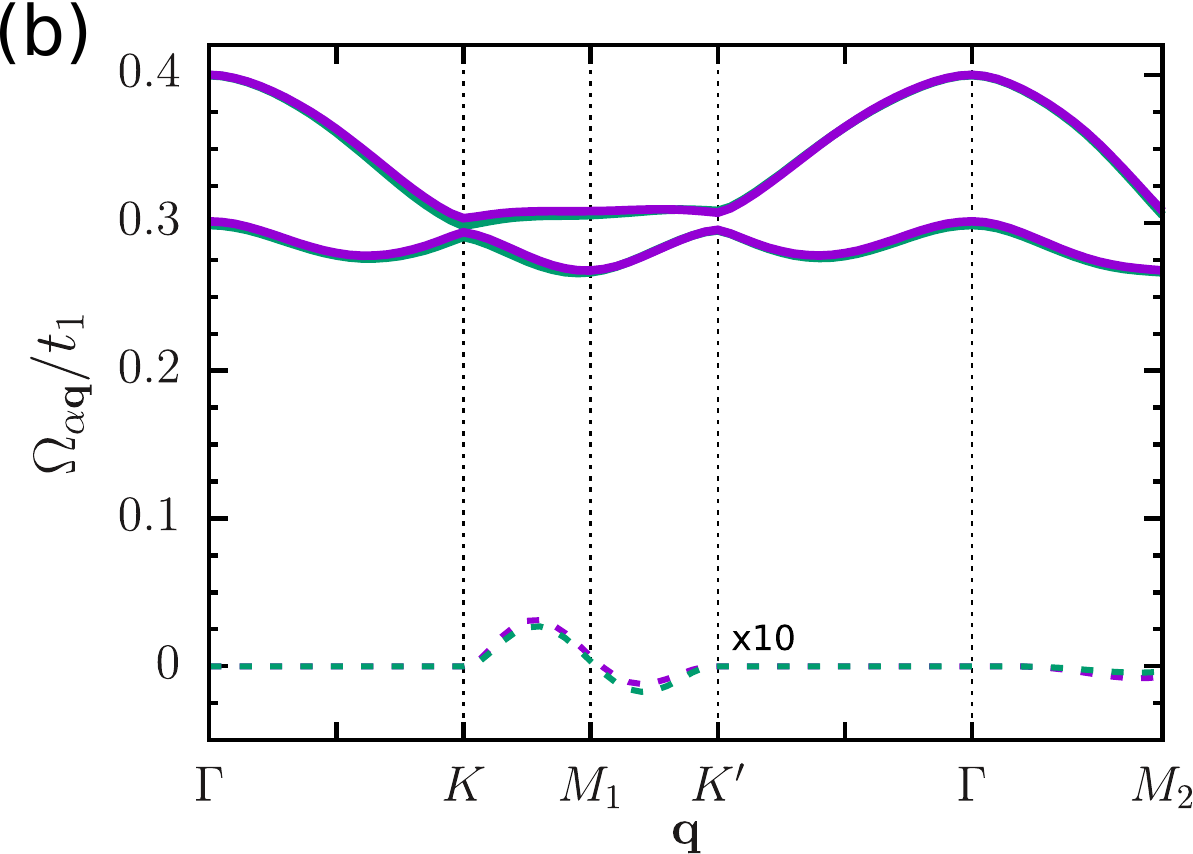}  
\hskip0.5cm
  \includegraphics[width=5.5cm]{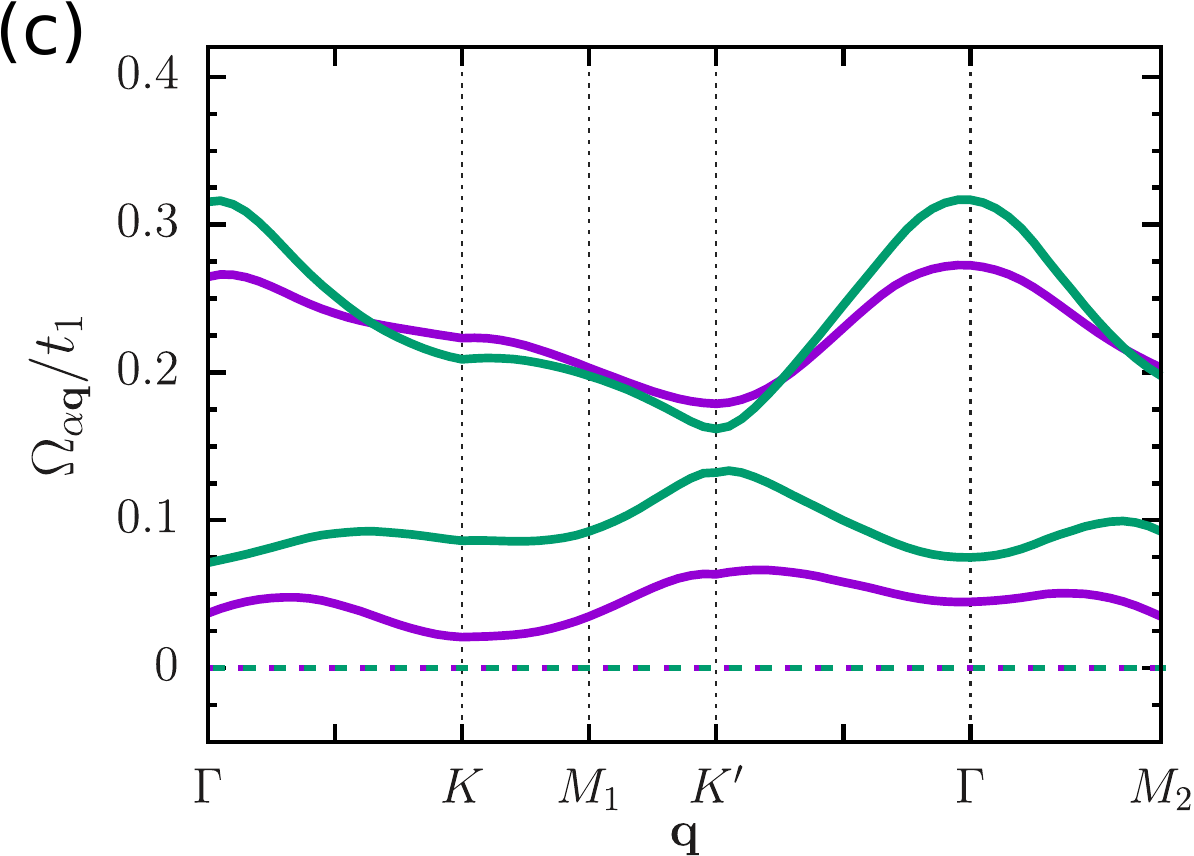}
} 
\caption{(a) Free electronic band structure \eqref{eq:omega} with the
additional staggered on-site energy term \eqref{eq:Hmass} 
along paths in the first BZ [Fig.~\ref{fig:Lattice}(b)]
for the nearly flat band limit \eqref{optimal-par} and staggered
on-site energy $M = 0.1\, t_1$: 
$\sigma = \uparrow$ (magenta) and
$\sigma = \downarrow$ (green).
(b) ML excitations \eqref{eq:projS1}: 
Spin-wave spectrum \eqref{omega-b}
along paths in the first BZ   
for the nearly flat band limit \eqref{optimal-par},
on-site Hubbard repulsion energies $U_A = U_B = U$,
and staggered on-site energy
$M=0.05$ (green) and 
$M=0.1\, t_1$ (magenta);
solid and dashed lines respectively represent the real part of $\Omega_{\pm,\bq}$
and the imaginary part of $\Omega_{+,\bq} = -\Omega_{-,\bq} $, where
the latter is multiplied by a factor of 10 for clarity.  
(c) Similar to panel (b), but for the SL excitations
\eqref{eq:projS2}.}
\label{fig:espectro-phi-M}
\end{figure*}

The spin-wave spectra \eqref{omega-b} 
for the nearly flat-band limit \eqref{optimal-par} and on-site repulsion energies 
$U_B = 0.8\, U_A = 0.8\, U$ and  $U_B = 0.6\, U_A = 0.6\, U$ 
are shown in Figs.~\ref{figEspectro}(b) and (c), respectively.
One sees that, as the diference $\Delta U = U_A - U_B$ increases: 
The energies of the spin-wave excitations decrease; 
the energy gap between the lower and upper bands at the $K$ point
increases,
\begin{align}
    \Delta^{(K)} &= 1.95 \times 10^{-2}\, U
    \quad {\rm for} \quad \Delta U = 0.2\,U,
\nonumber \\
    \Delta^{(K)} &= 2.85 \times 10^{-2}\, U
    \quad {\rm for} \quad \Delta U = 0.4\,U,
\nonumber
\end{align}
while the energy gap at the $K'$ point also varies,
\begin{align}
    \Delta^{(K')} &= 6.97 \times 10^{-4}\, U
    \quad {\rm for} \quad \Delta U = 0.2\,U,
\nonumber \\
    \Delta^{(K')} &= 1.12 \times 10^{-2}\, U
    \quad {\rm for} \quad \Delta U = 0.4\,U.
\nonumber 
\end{align}
For  $U_B > U_A$ (not shown here), similar features are observed, but now the energy
gap at the $K'$ point increases instead of the one at $K$ point with
the same overall intensities.  
Similarly to the homogeneous configuration $U_A = U_B  = U$, 
the spin-wave excitations \eqref{omega-b} at the border of the first BZ 
also have finite decay rates, which decrease as the diference
$\Delta U = U_A - U_B$ increases. 

The behaviour of the spin-wave spectrum \eqref{omega-b}
when the THM \eqref{eqHH} is moved away from
the nearly flat band limit \eqref{optimal-par} was also considered.
One calculates the spin-wave spectrum \eqref{omega-b} for 
$\phi = 0.4$, $0.5$, $0.7$, and $0.8$, 
hopping amplitude $t_2$ given by the relation $\cos(\phi) = t_1/(4 t_2)$, 
and homogeneous on-site repulsion energies $U_A = U_B = U$ (not shown here).
Similarly to the SL excitations [Figs.~\ref{fig:espectro-phi}(a) and (b)], 
one finds that the spin-wave spectra are
quite similar to the one obtained for the nearly-flat band limit
\eqref{optimal-par},  $\phi = 0.656$.

Even though the ML excitations are not the lowest-energy ones
for the correlated topological insulator \eqref{eqHH}, 
it would be interesting to see whether the
linear combinations \eqref{eq:projS1}  and \eqref{eq:projS2} 
could be slightly modified 
(e.g., with momentum dependent coefficients)
such that the condition
\eqref{conditionF} is now satisfied by all momenta in the first BZ.  
Such an modification may yield an Hermitian effective
boson model not only for the ML excitations, but also for the SL
excitations of the correlated Chern insulator \cite{leite2021}.
We left this issue for a future work.

%%%%%%%%%%%%%%%%%%%%%%%%%%%%%%%%%%%%%%%%%%%%%%%%%%%%%%%%%%%%%%%%%%%%%%%%%%%%%%%%%%%%%
\section{Staggered on-site energy term}
\label{sec:mass-term}

Here we briefly comment on the effects on the spin-wave spectrum \eqref{omega-b} 
due to the presence of an staggered on-site energy term, 
\begin{equation}
  H_M = \sum_{i \sigma} M \left( c_{i A \sigma}^{\dagger} c_{i A \sigma}   -  c_{i B \sigma}^{\dagger} c_{i B \sigma} \right),
\label{eq:Hmass}
\end{equation}
which breaks inversion symmetry when added to the noninteracting model
\eqref{eqHH0}.  
In the presence of the term \eqref{eq:Hmass}, it is easy to see that
the Hamiltonian \eqref{eqHH0} also assumes the form
\eqref{eq:Hfree}, with the dispersion of the free-electronic bands
given by  Eq.~\eqref{eq:omega} apart from the modification 
\begin{equation}
   B^\sigma_{3, \bk}  \to  B^\sigma_{3, \bk} + M 
                             = \gamma_\sigma B_{3,\bk} + M.
\label{eq:B3M}
\end{equation}
As discussed in detail in Sec.~II.B from Ref.~\cite{leite2021}, a
finite on-site energy $M$ increases the bandwidth of the lower
free-electronic band $c$, i.e., it allows us to move away from the
nearly flat-band limit, keeping the optimal parameter choice
\eqref{optimal-par} for $t_2$ and $\phi$.
Distinct from the Chern insulator on the hexagonal lattice
\cite{leite2021}, the staggered on-site energy term \eqref{eq:Hmass}
breaks the symmetry between
the spin $\uparrow$ and the spin $\downarrow$ free-electronic bands as
illustrated in Fig.~\ref{fig:espectro-phi-M}(a)
[note the $\gamma_\sigma$ factor in Eq.~\eqref{eq:B3M}].

Figures~\ref{fig:espectro-phi-M}(b) and (c) show the spin-wave
spectrum \eqref{omega-b} for $t_2$ and $\phi$ given by the optimal
parameter choice \eqref{optimal-par},
staggered on-site energy $M=0.05$ and $0.1\, t_1$, and
on-site repulsion energies $U_A = U_B = U = t_1$. 
For the ML excitations \eqref{eq:projS1} [Fig.~\ref{fig:espectro-phi-M}(b)],
a finite $M=0.05\, t_1$ yields minor
effects on the spin-wave spectrum as compared with the homogenous case
$M=0$ [Fig.~\ref{figEspectro}(a)].
Even for $M=0.1\, t_1$, the effects remain small, with
just a decreasing of the spin-wave energies around the $K$
point and an increasing in the energy gap between the lower and the
upper bands at the $K'$ point. 
Also, the decay rates  (the imaginary part of $\Omega_{\pm,\bq}$) 
display quite little modifications due to a finite  
$M$. On the other hand, for the SL excitations \eqref{eq:projS2}, the
effects related with a finite 
$M$ are more pronounced; see Fig.~\ref{fig:espectro-phi-M}(c).
Comparing with the homogeneous case $M=0$ [Fig.~\ref{figEspectro}(d)],
one notices that, as $M$ increases:
The energy gap between the lower and upper bands increases at the $K$
point and it has a nonmonotonic behavior at the $K'$ point;
the excitation gap of the lower band decreases and it moves from 
the $\Gamma$ point to the $K$ one. 
Such effects are qualitatively similar to the ones found  for on-site
repulsion energies $U_A \not= U_B$;
see Figs.~\ref{figEspectro}(e) and (f).
Interestingly, for $M=0.1\, t_1$, the excitation energy almost vanishes
at the $K$ point, a feature that could indicate an instability of
the flat-band FM phase. 
One should mention that, for the correlated Chern insulator
\cite{leite2021}, an instability of the flat-band FM phase
was found for any finite $M$. Finally, one should point out that, for
$M < 0$ (not shown here), the modifications in the 
spin-wave spectrum in the vicinity of the $K$ and $K'$ points are reversed.

The fact that the energy of the excitation gap monotonically decreases as
$M$ increases, as found for the SL excitations,
was previously observed for a time-reversal symmetric
THM on a square lattice \cite{neupert2012topological}.

%%%%%%%%%%%%%%%%%%%%%%%%%%%%%%%%%%%%%%%%%%%%%%%%%%%%%%%%%%%%%%%%%%%%%%%%%%%%%%%%%%%%% 
\section{Chern numbers of the spin-wave bands}
\label{sec:chernNumber2}

In this Appendix, we briefly describe the procedure employed to
numerically calculate the Chern numbers of the spin-wave bands
\eqref{omega-b}.  

We start casting the effective quadratic boson model
\eqref{H42} in a matrix form as done in Sec.\ref{sec:Diagonalization}
for the noninteracting Hamiltonian \eqref{eqHH0}, 
\begin{equation}
  \bar{H}_{U,B}^{(2)} = \sum_\bq  \Phi_\bq^\dagger  \tilde{h}_\bq \Phi_\bq ,  
\label{matrix-H42}
\end{equation}	
where the two-component spinor 
$\Phi_\bk = \left( b_{0, \mathbf{q}} \;\; 
			    b_{1, \mathbf{q}} \right)^T$, 
and the $2 \times 2$ matrix $\tilde{h}_\bq$ assumes the form
\begin{equation}
     \tilde{h}_\bq = \tilde{B}_{0, \bq }\tau_0 + \sum_{\mu=1}^3\tilde{B}_{\mu,\bq } \tau_\mu.
\end{equation}
Here $\tau_0$ is the identity matrix, $\tau_\mu$ is a Pauli matrix, and
\begin{align}
\tilde{B}_{0 , \bq} &= \frac{1}{2} \left( \epsilon_{\bq}^{00}  +\epsilon_{\bq}^{11}  \right),  
  \quad \quad 
&\tilde{B}_{1 , \bq} = \frac{1}{2}\left( \epsilon_{\bq}^{01} +\epsilon_{\bq}^{10} \right),  
\nonumber\\
\tilde{B}_{2 , \bq} &= \frac{1}{2i} \left( \epsilon_{\bq}^{01}  -\epsilon_{\bq}^{10}  \right), 
&\tilde{B}_{3 , \bq} = \frac{1}{2} \left( \epsilon_{\bq}^{00} -\epsilon_{\bq}^{11} \right),  
\label{coef-Btilde}
\end{align} 
with $\epsilon_{\bq}^{\alpha\beta}$ being the coefficients \eqref{eq:Epsilon}. 

\begin{figure}[t]
\centerline{\includegraphics[width=6.1cm]{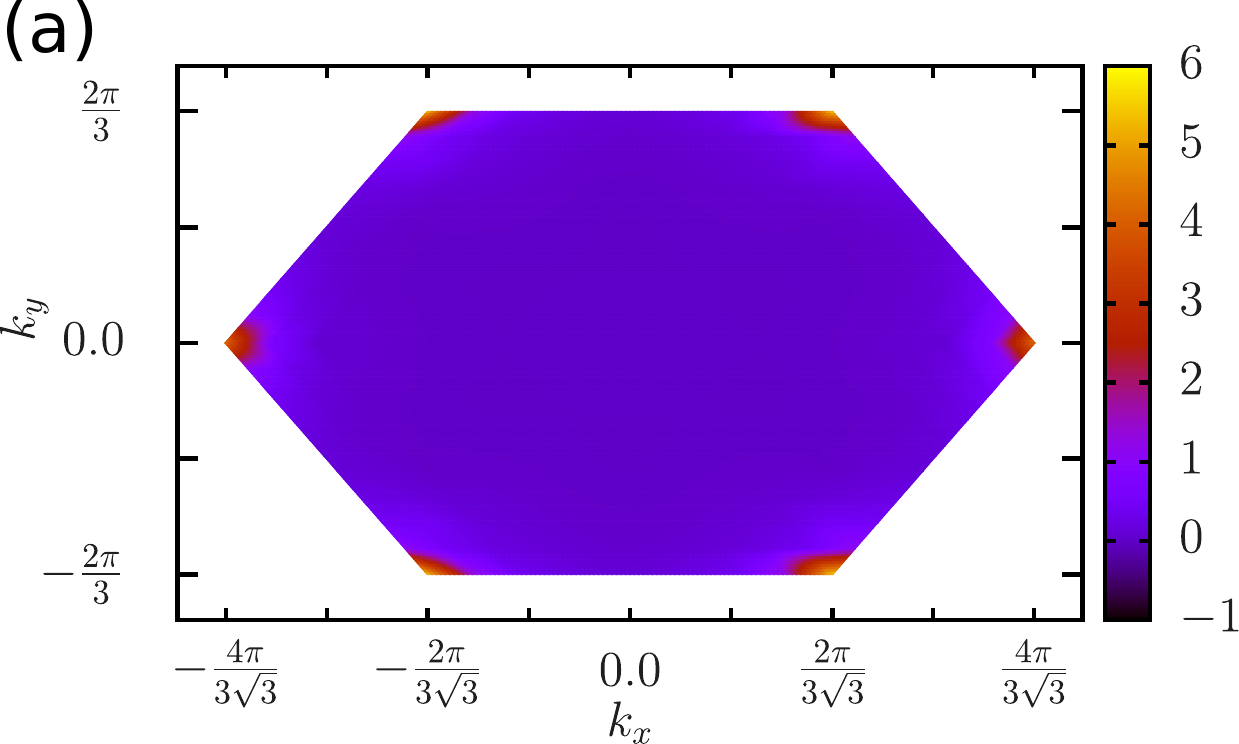}} 
\vskip0.4cm
\centerline{\includegraphics[width=6.1cm]{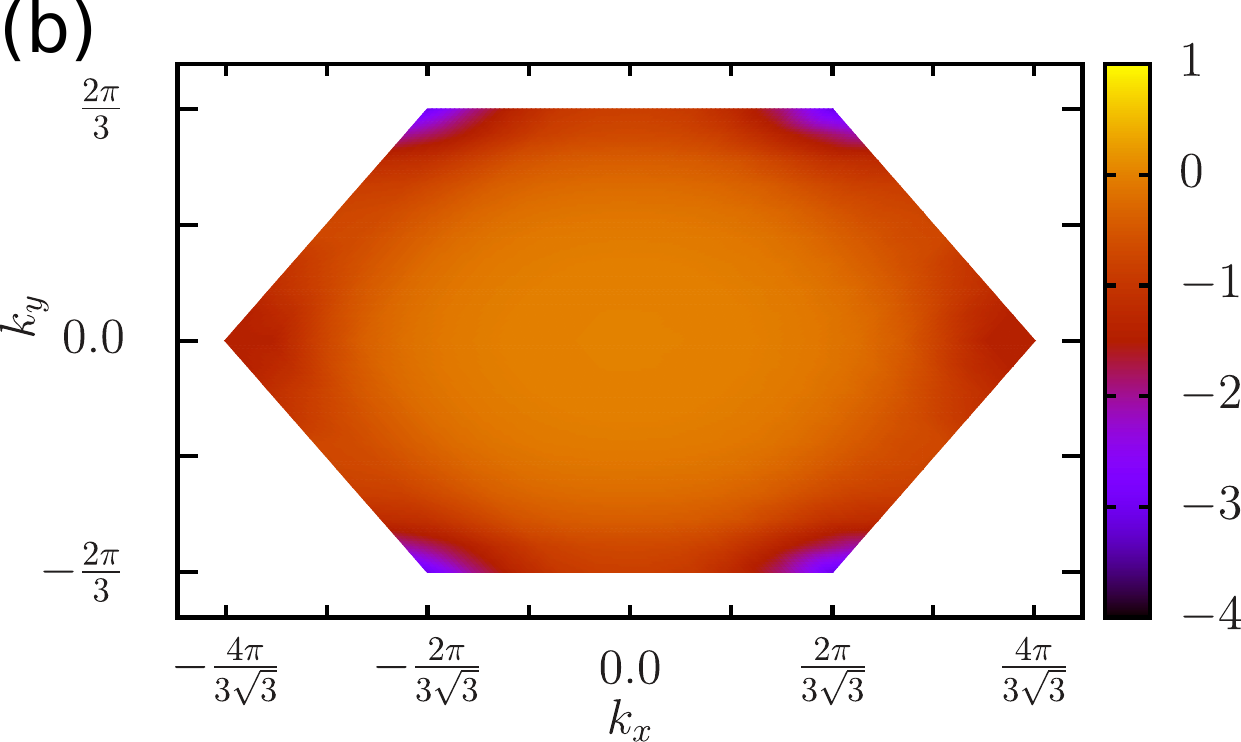}}
\caption{Contour plot of the Berry curvature of the lower spin-wave
  band within the first BZ for the nearly flat band limit
  \eqref{optimal-par} and on-site 
  repulsion energies $U_A = U_B = U$.
  (a) ML [Fig.~\ref{figEspectro}(a)] and   
  (b) SL [Fig.~\ref{figEspectro}(d)] excitations.}  
\label{fig:curvatures}
\end{figure}

Due to the similarities between the forms of the Hamiltonians
\eqref{eqH0k} and \eqref{matrix-H42},
the Chern numbers of the spin-wave bands \eqref{omega-b}
are also given by Eq.~\eqref{eqCn}, apart from the $\gamma_\sigma$
factor and the replacement 
$\hat{B}_{\mu,\bk} \to \tilde{B}_{\mu,\bk} / |\tilde{\textbf{B}}_\bk|$,
where
$ |\tilde{\textbf{B}}_\bk| = \sqrt{ \tilde{B}^2_{1,\bk} +
                                                    \tilde{B}^2_{2,\bk} + \tilde{B}^2_{3,\bk} }$.
Moreover, for the ML excitations, we assume that 
$\epsilon^{01}_\bq = (\epsilon^{10}_\bq)^*$ in order to obtain real Chern numbers:
Recall that, only for the ML excitations, the quadratic Hamiltonian
\eqref{H42} is non-Hermitian, see Appendix~\ref{sec:ap-spin-wave};
such an assumption was also made in Ref.~\cite{leite2021} in order to
determine the Chern numbers of the spin-wave bands of a correlated
Chern insulator.

The Berry curvature, which is defined as one-half of the integrand of 
Eq.~\eqref{eqCn}, of the lower spin-wave band \eqref{omega-b} 
for the nearly flat-band limit \eqref{optimal-par} and on-site
repulsion energies $U_A = U_B = U$ is shown in Fig.~\ref{fig:curvatures}. 
For both ML and SL excitations, one sees that the Berry curvatures
peak at the $K$ and $K'$ points of the first BZ.

The Chern numbers of the lower spin-wave bands \eqref{omega-b} 
for both the ML ($C_{ML}$) and the SL ($C_{SL}$) excitations, 
which are determined by numerically
integrating Eq.~\eqref{eqCn} and considering the coefficients
\eqref{coef-Btilde}, are shown in Table~\ref{tb-chern}. 
For the SL excitations shown in  Figs.~\ref{figEspectro}(d) and (e),
one sees that the Chern numbers are close to one. We believe that the small deviations
from the unit might be due to the fact that the numerical
procedure used to calculated the Chern numbers does not properly take
into account the behaviour of the Berry curvature [Fig.~\ref{fig:curvatures}(b)] 
at the corners of the first BZ.  
On the other hand, for the ML excitations shown in Figs.~\ref{figEspectro}(a) and (b),
the Chern numbers are finite, but smaller than one. In addition to
possible numerical issues [see Fig.~\ref{fig:curvatures}(a)],  
such fractional values for the Chern numbers might also be associated to
the fact that it is necessary to assume that    
$\epsilon^{01}_\bq = (\epsilon^{10}_\bq)^*$ in order to obtain real Chern numbers.

The nonzero Chern numbers found for the spin-wave bands of the
correlated topological insulator are in constrast with the topological
properties of the corresponding correlated Chern insulator on a
honeycomb lattice \cite{gu2019itinerant,leite2021}.
In the completely flat band limit, i.e, when the dispersion of the
noninteracting electronic bands is neglected (an
approximation similar to the assumption $\bar{\omega}^{\alpha\beta}_\bq = 0$
made in Appendix~\ref{sec:ap-spin-wave}
and in Ref.~\cite{leite2021} that the kinetic coefficients
\eqref{eq:omegaBar} vanish), 
it was found that the spin-wave bands of the correlated Chern
insulator are topologically trivial \cite{gu2019itinerant,leite2021}.   
Indeed, our previous results \cite{leite2021} are in agreement with
the exact diagonalization calculations \cite{gu2019itinerant}.
Moreover, it was also numerically shown \cite{gu2019itinerant} 
that the spin-wave bands of the correlated Chern insulator acquire
nonzero Chern numbers when the 
dispersion of the free-electronic bands is explicitly taken into
account (see also Sec.~V from Ref.~\cite{leite2021}).

Although it is not clear whether the spin-wave bands for the ML
excitations \eqref{eq:projS1} are topologically nontrivial,
one finds some evidences that
the spin-wave bands for the SL excitations \eqref{eq:projS2} might be
topologically nontrivial, even in the completely flat band limit of the
free-electronic bands, a feature that contrasts with the
behaviour of the corresponding correlated Chern insulator.

%%%%%%%%%%%%%%%%%%%%%%%%%%%%%%%%%%%%%%%%%%%%%%%%%%%%%%%%%%%%%%%%%%%%%%%%%%%%%%%%%%%%

\end{document}